%% file: paper.tex
\documentclass[12pt]{article}
\usepackage{graphics}

\newcommand{\MS}{\overline{\rm MS}}
\newcommand{\be}{\begin{equation}}
\newcommand{\ee}{\end{equation}}
\newcommand{\bea}{\begin{eqnarray}}
\newcommand{\eea}{\end{eqnarray}}
\newcommand{\lra}{\leftrightarrow}
\newcommand{\LL}{\mathcal{L}}

\newcommand{\psibar}{\overline{\psi}}
\newcommand{\cbar}{\overline{c}}
\newcommand{\hD}[1]{\stackrel{\leftrightarrow}{D^{#1}}}
\newcommand{\shD}{\stackrel{\leftrightarrow}{D\!\!\!\! / \,}}
\newcommand{\slsh}[1]{{#1}\!\!\! / \,}
\newcommand{\dx}{{\rm d}^3x}
\newcommand{\mat}[3]{\left<{#1}\left|{#2}\right|{#3}\right>}
\newcommand{\ket}[1]{\left.\left|{#1}\right.\right>}
\newcommand{\pp}{p^\prime}
\newcommand{\derd}[2]{\frac{\partial}{\partial{#1}_{#2}}}

\newcommand{\meas}[2]{\frac{d^{#2}{#1}}{(2\pi)^{#2}}}
\newcommand{\coupling}{\frac{g^2}{(4\pi)^2}}
\newcommand{\pole}{\frac{1}{\epsilon}}

\newlength{\updownindent}
\newlength{\leftrightindent}
\begin{document}

\thispagestyle{empty}
\renewcommand{\thefootnote}{\alph{footnote}}

\begin{flushright} CERN-TH/99-390\\ SWAT/99-248 \\
  UGVA-DPT-99-12-1067 \end{flushright}

\begin{center}
  {\Large \bf The Gauge-Invariant Angular Momentum Sum-Rule for the Proton}\\
\vspace{0.2in}
  {\bf G. M. Shore}\footnote{%
    Permanent address: Physics Department, University of Wales Swansea,
    Singleton Park, Swansea SA2 8PP, U.K.}\\
\vspace{0.1in}
  {\it D\'epartement de Physique Th\'eorique, Universit\'e de Gen\`eve,}\\
  {\it 24, quai E. Ansermet, CH-1211 Gen\`eve 4, Switzerland,}\\
    and\\
  {\it TH Division, CERN,} \\
  {\it CH-1211 Gen\`eve 23, Switzerland. }\\
  E-mail: {\tt g.m.shore@swansea.ac.uk}\\
\vspace{0.2in}
  {\bf B. E. White}\\
\vspace{0.1in}
  {\it Physics Department, University of Wales Swansea,} \\
  {\it Singleton Park, Swansea SA2~8PP, UK.}\\
  E-mail: {\tt pywhite@swan.ac.uk}\\
\vspace{0.15in}
\end{center}

\begin{abstract}

We give a gauge-invariant treatment of the angular momentum sum-rule for
the proton in terms of matrix elements of three gauge-invariant, local
composite operators. These matrix elements are decomposed into three
independent form factors, one of which is the flavour singlet axial
charge. The other two are interpreted as total quark and gluon
angular momentum.
We further show that the axial charge cancels out of the sum-rule.
The general form of the renormalisation mixing of
the three operators is written down and also determined to one
loop from which the scale dependence and mixing
of the form factors is derived. We relate these results to a previous
parton model calculation by defining the parton model quantities in terms
of the three form factors.
We also mention how the form factors can be measured in experiments.

\end{abstract}

\vspace{0.3in}
\begin{flushleft}CERN-TH/99-390 \\ December 99
\end{flushleft}

\newpage
\setcounter{page}{1}

\renewcommand{\thefootnote}{\arabic{footnote}}
\setcounter{footnote}{0}


\section{Introduction}

In the simplest parton model of the proton, its spin is carried
by the spin of the valence quarks in such a way that their overall spin
state is that of a spin-1/2 particle. In a QCD-improved parton model,
however, such a picture must evolve with the renormalisation scale due to the
splitting processes $q \rightarrow q g$ etc;
at an arbitrary scale, some of the proton spin is then in principle carried
by gluons and by strange quarks etc. Furthermore, it was realised some
time ago by Ratcliffe~\cite{Rat}
that the parton splitting processes generate orbital angular
momentum. Thus, if the total
spin of the proton is to have a scale-invariant meaning
in the context of a parton model, then we must introduce quark and gluon
orbital angular momentum components $L_q(x)$ and $L_g(x)$. Together with
the usual quark and gluon intrinsic spin components $\Delta q(x)$
and $\Delta g(x)$, which are extracted from
polarised Deep Inelastic Scattering (DIS), the sum of their first moments
should be scale-invariant.

The splitting functions for these four distributions were calculated by
Hoodbhoy, Ji and Tang (HJT)~\cite{HJT} to leading order. The form of the mixing
of the first moments\footnote{%
Higher moments were calculated in refs.~\cite{HS,HK}.}
under scaling was found to be more general
than originally envisaged by
Ratcliffe~\cite{Rat}. Importantly, splitting functions do not depend
on the choice of gauge used in their calculation. However, it is also claimed
by HJT that the first moments of the angular momentum components, $\Delta q$,
$\Delta g$, $L_q$ and $L_g$, can be represented as forward matrix elements
of four composite operators in proton states. Only one of these operators is
gauge-invariant. This is the source of many difficulties, some of which are
reported in a later paper~\cite{HJT2}.
As an alternative, if we would define the four components of the proton
spin in terms of forward matrix elements of gauge-invariant composite
operators, we have a different challenge --- there exists only one
gauge-invariant operator which can measure total
gluon angular momentum ($\Delta g$ and $L_g$ in the parton model). This means
one has to relate four parton model quantities to just three operators. This
is discussed in refs.~\cite{JM,Ji}. Moreover, as far as we know, there
exists no operator definition of $L_q(x)$ and $L_g(x)$ which scale in the
manner of HJT.

In this paper, we study quantities which have a precise
operator definition. We discuss the gauge-invariant decomposition
of the angular momentum current into three operators, thus restricting
ourselves to quantities which can be interpreted as first moments only.
We show that the forward matrix
elements of the three gauge-invariant operators can be decomposed into just
three Lorentz-scalar form factors. One of these form factors is the proton's axial charge which is measured from the flavour singlet component of the first moment of $g_1$ in polarised DIS. Remarkably, we will show that the axial charge cancels out of the angular momentum sum-rule. We believe that this point
has not been emphasised before.
Since only two form factors appear in the angular momentum
sum rule, we must conclude that the proton spin is only meaningfully decomposed
into two pieces.
We will interpret the two form factors as total quark and gluon angular
momentum. Further, we need not interpret the axial charge as a measure of
intrinsic quark spin. Indeed, because of the axial anomaly, the axial charge of the proton is, in the chiral limit, equal to its topological charge,
which does not have an obvious interpretation as quark spin since it is
defined as the matrix element of a gluon operator. An independent treatment of the anomalous suppression of the first moment of $g_1$ (the ``proton spin'' problem) can thus be given in terms of universal topological charge screening~\cite{NSV}.

We also derive
the renormalisation mixing of the three gauge-invariant operators.
We show how, in terms of operator vertices, the axial anomaly affects two of 
the operators (one is the axial current) so that their sum is free of the
anomaly. This is consistent with the fact that the axial charge does not appear
in the sum-rule in terms of form factors. We have performed a one-loop
calculation of the mixing of the operators.

After presenting our calculations, we then show how they
are related to the parton model calculation of HJT. To do this, we shall need
to make an identification between the three form factors including the axial
charge and the four parton model quantities. This is necessarily a Lorentz
frame dependent statement, but which reproduces their results.

Recently, it was proposed~\cite{Ji,Ji2}
that the two new form factors could be measured in hard, exclusive processes
$\gamma^\ast p \rightarrow X p$ such as Deeply Virtual Compton
Scattering~(DVCS) where $X = \gamma$. A section reviewing this possibility
is included.

The paper is organised as follows:
In section~\ref{sec_decomp} we review the decomposition of the conserved
angular momentum current (AMC) into spin and orbital pieces, explaining
why the gluon piece cannot be decomposed in this way gauge-invariantly.
We include the gauge-fixing terms from the Lagrangian. Our
main results are contained in sections~\ref{sec_role} and~\ref{sec_leading}: In
section~\ref{sec_role} we decompose the forward matrix elements of the three
operators in the gauge-invariant decomposition of the AMC into scalar form
factors; section~\ref{sec_leading} contains the
results of a one-loop
calculation of the divergent pieces of the three operators, from which are
derived the evolution and mixing of the form factors. In
section~\ref{sec_parton} we discuss how to relate the form factors to parton
model quantities, and how the result of HJT is obtained.
In section~\ref{sec_meas}
we review how the two new form factors can be extracted from a class
of hard, exclusive processes such as Deeply Virtual Compton Scattering (DVCS).
Finally, section~\ref{sec_gninv} reviews attempts made in the literature
to identify angular momentum components of the proton to matrix elements
of non-gauge-invariant operators, highlighting the problems with such
treatments.


\section{Decomposition of Angular Momentum Current}
\label{sec_decomp}

In this section, we summarise the construction of a conserved angular
momentum current (AMC)
$M^{\mu\nu\lambda}$ from the QCD lagrangian. As with the
stress-energy tensor $T^{\mu\nu}$, we have the freedom to redefine
$M^{\mu\nu\lambda}$ by adding on divergence pieces such that
$M^{\mu\nu\lambda}$ is still conserved.
As is discussed in ref.~\cite{JM},
it is possible in this way to separate the quark part of $M^{\mu\nu\lambda}$
into spin and orbital angular momentum (OAM)
pieces. However, it is not possible
to do the same for the gluon term in a gauge-invariant fashion. The reason
is that the gauge field's spin and space-time indices coincide. Below,
we extend the discussion in ref.~\cite{JM} to include extra terms coming
from gauge-fixing terms in covariant gauges and show that
the decomposition of the AMC will depend on the details of the gauge-fixing
procedure, so that even the ghost fields carry OAM. We will conclude from this
that it is not sensible to define observables except in terms of the gauge
invariant decomposition, even though this means we are left with only three
operators measuring observable quantities instead of four.

We begin with the divergence-free, symmetric stress-energy tensor $T^{\mu\nu}$.
From it we can construct $M^{\mu\nu\lambda}$:
\be
  M^{\mu\nu\lambda} = x^{[\nu} T^{\lambda]\mu},
\ee
where the square brackets denote antisymmetrisation.\footnote{%
Our convention is that $a^{[\mu}b^{\nu]} = a^\mu b^\nu - a^\nu b^\mu$, while
$a^{\{\mu}b^{\nu\}} = a^\mu b^\nu + a^\nu b^\mu$.}
This is conserved
($\partial_\mu M^{\mu\nu\lambda} = 0$) provided that $T^{\mu\nu}$ is symmetric.
It is always possible to redefine $M^{\mu\nu\lambda}$ by adding the divergence
of a function $X^{\mu\alpha\nu\lambda}$:
\be
  M^{\prime \, \mu\nu\lambda} = M^{\mu\nu\lambda} + \partial_\alpha
  X^{\mu\alpha\nu\lambda}.
\ee
$M^\prime$ is also conserved provided $X$ is antisymmetric under
$\mu \lra \alpha$ as well as under $\nu \lra \lambda$.\footnote{%
Note that a transformation $T^{\prime \, \mu\nu} = T^{\mu\nu} + \partial_\alpha
X^{\mu\nu\alpha}$ does not correspond to a transformation of
$M^{\mu\nu\lambda}$.}

If we now consider the covariantly gauge-fixed, BRST-invariant QCD lagrangian,
\bea
  \LL & = & \LL_{\rm gi} + \LL_{\rm gf},\nonumber \\
  \LL_{\rm gi} & = & \psibar i\shD \psi - \frac{1}{4} F^{\mu\nu}F_{\mu\nu},\\
  \LL_{\rm gf} & = & -i \partial^\mu \cbar D_\mu c - \partial^\mu B A_\mu
   + \frac{\alpha}{2} B^2,\nonumber
\eea
the corresponding symmetric stress-energy tensor is
\bea
  T^{\mu\nu} & = & T^{\mu\nu}_{\rm gi} + T^{\mu\nu}_{\rm gf}, \nonumber \\
  T^{\mu\nu}_{\rm gi} & = & \frac{1}{2}\psibar \gamma^{\{\mu} i\hD{\nu\}} \psi
   + F^{\mu\alpha}F_\alpha^{\phantom{a}\nu} - g^{\mu\nu} \LL_{\rm gi},\\
  T^{\mu\nu}_{\rm gf} & = & -i\partial^{\{\mu}\cbar D^{\nu\}}c
   - \partial^{\{\mu}B A^{\nu\}} - g^{\mu\nu}\LL_{\rm gf}.\nonumber
\eea
Writing $M^{\mu\nu\lambda}_{\rm gi} = x^{[\nu} T^{\lambda]\mu}_{\rm gi}$, one
can show that
\bea
   M^{\mu\nu\lambda}_{\rm gi} & = & \frac{1}{2} \epsilon^{\mu\nu\lambda\alpha}
    \psibar \gamma_\alpha \gamma^5 \psi + \psibar \gamma^\mu x^{[\nu}
    i\hD{\lambda]}\psi + x^{[\nu} F^{\lambda]\alpha}
    F_\alpha^{\phantom{a}\mu} - x^{[\nu} g^{\lambda]\mu} \LL_{\rm gi}
    \nonumber \\
   & - & \frac{1}{4} \partial_\alpha \left[ x^{[\nu}
    \epsilon^{\lambda]\mu\alpha\beta} \psibar \gamma_\beta \gamma^5 \psi
    \right] + {\rm EOM}.
\label{gi_sep}\eea
Defining $M^\prime$ so as to cancel the divergence and equation of motion~(EOM)
pieces, we have
\bea
  M^{\prime \, \mu\nu\lambda} & = & M^{\prime \, \mu\nu\lambda}_{\rm gi}
   + M^{\prime \, \mu\nu\lambda}_{\rm gf},
   \nonumber \\
  M^{\prime \, \mu\nu\lambda}_{\rm gi} & = & O_1^{\mu\nu\lambda}
   + O_2^{\mu[\lambda} x^{\nu]} + O_3^{\mu[\lambda} x^{\nu]}
   - x^{[\nu} g^{\lambda]\mu} \LL_{\rm gi}, \nonumber \\
  M^{\prime \, \mu\nu\lambda}_{\rm gf} & = & x^{[\nu} T^{\lambda]\mu}_{\rm gf}
\label{gi_decomp}\eea
with
\bea
  O_1^{\mu\nu\lambda} & = & \frac{1}{2} \epsilon^{\mu\nu\lambda\alpha}
   \psibar \gamma_\alpha \gamma^5 \psi,\nonumber \\
  O_2^{\mu\nu} & = & \psibar \gamma^\mu i\hD{\nu} \psi,\\
  O_3^{\mu\nu} & = & F^{\mu \alpha} F_\alpha^{\phantom{a}\nu}.\nonumber
\eea
We have in this way split the quark piece into what appears to be a spin piece, the axial current
$O_1$, and an OAM piece $xO_2$, while $xO_3$ is the total gluon angular
momentum, both spin and OAM. All the terms in $M^\prime_{\rm gi}$ are
gauge-invariant. Matrix elements of $M^\prime_{\rm gf}$ vanish since it is
a BRST variation of another operator. Also, forward matrix elements of the term
propertional to $x^{[\nu} g^{\lambda]\mu}$ vanish;
this will be shown in the next section. It is this decomposition which we
will use to derive the scaling of the components of a nucleon's spin: It
ensures that all quantities are explicitly gauge-invariant.

Now consider what we get when we try to split the gluon component $O_3$ into
spin and OAM parts. Again, one can show that
\bea
  M^{\prime \, \mu\nu\lambda} & = & \frac{1}{2} \epsilon^{\mu\nu\lambda\alpha}
   \psibar \gamma_\alpha \gamma^5 \psi +
   i\psibar \gamma^\mu x^{[\nu} \partial^{\lambda]}\psi \nonumber \\
  & - & F^{\mu[\nu} A^{\lambda]} -
   F^{\mu\alpha} x^{[\nu}\partial^{\lambda]} A_\alpha \nonumber \\
  & - & A^\mu x^{[\nu}\partial^{\lambda]} B
   - i\partial^\mu\cbar x^{[\nu}\partial^{\lambda]}c
   - i x^{[\nu}\partial^{\lambda]} \cbar D^\mu c - x^{[\nu} g^{\lambda]\mu}
   \LL\nonumber \\
  & + & \partial_\alpha \left[ F^{\mu\alpha} x^{[\nu} A^{\lambda]}\right]
   + {\rm EOM}.
\label{nongi_decomp}\eea
We can redefine $M^\prime \rightarrow M^{\prime\prime}$ again by dropping
the divergence and EOM terms. The new decomposition coincides with that of
the canonical angular momentum current: Defining the rotation generator
$J^{ij}$,
\be
  J^{ij} = \int (\dx)_\mu M^{\mu ij},
\label{spin_vector}\ee
if we choose the integral to be over 3-dimensional space at a given time, we
have
\bea
  J^{ij}_{(0)} & = & \int \dx M^{0ij}\nonumber \\
  & = & \int \dx \left[ \psi^\dagger \sigma^{ij} \psi +
   i \psi^\dagger x^{[i}\partial^{j]} \psi
   - E^{[i} A^{j]} + E^k x^{[i}\partial^{j]} A^k
   \right.\nonumber\\
  & - & \left. A^0 x^{[i}\partial^{j]} B
   - i\partial^0\cbar x^{[i}\partial^{j]} c
   - ix^{[i}\partial^{j]}\cbar D^0 c\right],
\eea
with $E^i = F^{0i}$ and $\sigma^{ij} = \frac{i}{4}[\gamma^i,\gamma^j]$.
This may be written
\be
  J^{ij}_{(0)} = \sum_\phi \int \dx \; i\pi(\phi_b) (\delta^{ij})_{ba} \phi_a,
\label{canon}\ee
where $\phi_a$ are the canonical fields of the Kugo-Ojima canonical formulation
of non-Abelian gauge
theory~\cite{KO}, $\pi(\phi_a)$ are their conjugate momenta, and
$\delta^{ij}$ are their variations under a spatial rotation.\footnote{%
Each term in $J^{ij}_{(0)}$ satisfies
the generator algebra for rotations. Note,
however, that when the operators are renormalised, this is no longer true.
Indeed, this even happens for the first term, the axial current, due to the
axial anomaly.}

One might be tempted to take the first four terms in eq.~(\ref{nongi_decomp})
and identify their matrix elements with quark spin and OAM and gluon spin and
OAM respectively. There are good reasons for not doing so:
\begin{enumerate}
  \item The other terms in eq.~(\ref{nongi_decomp}) involving the gauge-fixing
fields $c, \cbar, B$ are no longer a BRST invariant combination;
the sum of their
matrix elements cannot be expected to vanish. This means that we would have to
talk about ghost OAM etc. in order to have a basis of operators which is
closed upon renormalisation.
  \item The mixing amongst quark and gluon spin and OAM will depend on the
gauge parameter, as observed in ref.~\cite{HJT}. The observables whose scaling
we wish to derive, on the other hand, are of course gauge invariant.
\end{enumerate}


\section{Role of the Axial Charge}
\label{sec_role}

We now develop further the gauge-invariant decomposition of the angular
momentum current (AMC) in eq.~(\ref{gi_decomp}). First, we decompose the forward matrix elements of the operators in the AMC into three Lorentz scalar form factors. One of these, the axial charge, appears twice with opposite sign, and so cancels out of the angular momentum sum-rule. This allows us to conclude that firstly the axial charge does not form part of the angular momentum sum rule
and secondly that we do not need to interpret the axial charge as a measure
of parton spin. The other two form factors have been discussed
before~\cite{JM,Ji}, however the relationship of these to the axial anomaly
has not been emphasised. Furthermore, if the axial charge is to decouple from
the sum rule in this way, we should be able to show this is reflected in the
renormalisation of the AMC. To our knowledge, an explicit derivation of the
mixing of the
three gauge invariant operators appearing in the AMC has not yet appeared.
Therefore, below we derive the general form of the mixing. In the next section
we calculate the mixing explicitly to one loop.

The axial charge $a_0$ is defined to be a Lorentz scalar
form factor, or reduced matrix element:
\be
  \mat{p,s}{O^{\mu\nu\lambda}_1}{p,s} =
   M\epsilon^{\mu\nu\lambda\rho} s_\rho a_0,
\label{axial}\ee
We should be able to define all our observables
in terms of such scalar functions, so let us now decompose forward matrix
elements of all the remaining operators in the AMC~(\ref{gi_decomp}) except
the divergence and the EOM, i.e.
$O_2^{\mu[\lambda}x^{\nu]}$, $O_3^{\mu[\lambda}x^{\nu]}$
and $g^{\mu[\lambda}x^{\nu]} \LL$.
To do this, we need to define the forward matrix element of an operator
involving a factor of the co-ordinate $x^\mu$. This is done in Appendix A.
(See refs.~\cite{HJT,HS}.)
The result is that if $B^{\mu\nu}(x)$ is a local composite operator, then the
forward matrix element of $x^{[i}B^{j]0}(x)$ is
\be
  \Phi^{-1}(p) \int \dx \mat{p}{x^{[i}B^{j]0}(x)}{\Phi}
   = \left. -i \derd{\Delta}{[i} \mat{p}{B^{j]0}(0)}{\pp} \right|_{\pp=p},
\label{for_mat}\ee
where $\Phi(p)$ is a wavepacket with no azimuthal dependence about the
k-direction and where $P=\frac{1}{2}(p+\pp)$ is held constant in the partial derivative w.r.t. $\Delta=p-\pp$.
Before using this expression, we decompose the off-forward matrix elements
of $g^{\mu\nu} \LL$, $O_2^{\mu\nu}$, $O_3^{\mu\nu}$ and
\be
  O_4^{\mu\nu} = \frac{1}{2} \psibar \gamma^{\{\mu}i\hD{\nu\}} \psi
\ee
into scalar form factors. We write down only those terms which are linear
in $\Delta$ and which are even in $P$; the
latter constraint comes from symmetry under crossing,
$\Delta \rightarrow \Delta$ and $P \rightarrow -P$. The Lorentz structures
must also have even parity:
\bea
  \mat{p,s}{g^{\mu\nu}\LL}{\pp,s} & = & C_\LL(\Delta^2) M^2 g^{\mu\nu},
   \nonumber \\
  \mat{p,s}{O_2^{\mu\nu}(0)}{\pp,s} & = & A_q(\Delta^2)P^\mu P^\nu
   + \frac{B_q(\Delta^2)}{2M} P^{\{\mu}\epsilon^{\nu\}\alpha\beta\sigma}
   i\Delta_\alpha P_\beta s_\sigma + C_q(\Delta^2) M^2 g^{\mu\nu}
   \nonumber \\
  & + & \frac{\widetilde{B}_q(\Delta^2)}{2M}
   P^{[\mu}\epsilon^{\nu]\alpha\beta\sigma} i\Delta_\alpha P_\beta s_\sigma
   + MD_q(\Delta^2)\epsilon^{\mu\nu\alpha\beta}i\Delta_\alpha s_\beta, \\
  \mat{p,s}{O_3^{\mu\nu}(0)}{\pp,s} & = & A_g(\Delta^2)P^\mu P^\nu
   + \frac{B_g(\Delta^2)}{2M} P^{\{\mu}\epsilon^{\nu\}\alpha\beta\sigma}
   i\Delta_\alpha P_\beta s_\sigma + C_g(\Delta^2) M^2 g^{\mu\nu},\nonumber\\
  \mat{p,s}{O_4^{\mu\nu}(0)}{\pp,s} & = & A_q(\Delta^2)P^\mu P^\nu
   + \frac{B_q(\Delta^2)}{2M} P^{\{\mu}\epsilon^{\nu\}\alpha\beta\sigma}
   i\Delta_\alpha P_\beta s_\sigma + C_q(\Delta^2) M^2 g^{\mu\nu}.\nonumber
\eea
Notice that the symmetric part of $O_4$ is identical to that of $O_2$.
Substituting these in the r.h.s. of~(\ref{for_mat}) we get
\bea
  \left. -  i\derd{\Delta}{\nu}\mat{p,s}{g^{\mu\lambda}\LL(0)}{\pp,s}\right|_{p=\pp}
   & = & 0, \nonumber \\
  \left. -  i\derd{\Delta}{\nu}\mat{p,s}{O_2^{\mu\lambda}(0)}{\pp,s}\right|_{p=\pp}
   & = & \frac{B_q(0)}{2M} p^{\{\mu}\epsilon^{\lambda\}\nu\beta\sigma}
   p_\beta s_\sigma\nonumber, \\
  & + & \frac{\widetilde{B}_q(0)}{2M} p^{[\mu}\epsilon^{\lambda]\nu\beta\sigma}
   p_\beta s_\sigma
   - MD_q(0)\epsilon^{\mu\nu\lambda\beta}s_\beta, \label{am_form}\\
  \left. -i\derd{\Delta}{\nu}\mat{p,s}{O_3^{\mu\lambda}(0)}{\pp,s}\right|_{p=\pp}
   & = & \frac{B_g(0)}{2M} p^{\{\mu}\epsilon^{\lambda\}\nu\beta\sigma}
   p_\beta s_\sigma,\nonumber\\
  \left. -i\derd{\Delta}{\nu}\mat{p,s}{O_4^{\mu\lambda}(0)}{\pp,s}\right|_{p=\pp}
   & = & \frac{B_q(0)}{2M} p^{\{\mu}\epsilon^{\lambda\}\nu\beta\sigma}
   p_\beta s_\sigma \nonumber.
\eea
There are further constraints on the five remaining form factors: We
showed in deriving eq.~(\ref{gi_sep}) that
\be
  O_4^{\mu[\lambda} x^{\nu]} = O_1^{\mu\nu\lambda} + O_2^{\mu[\lambda} x^{\nu]}
   + {\rm divergence} + {\rm EOM},
\label{split}\ee
It is shown in Appendix A that the forward matrix element of the divergence
term vanishes. It follows from eqs.~(\ref{axial},\ref{am_form},\ref{split})
that
\bea
  \widetilde{B}_q(0) & = & 0, \nonumber \\
  2 D_q(0) & = & a_0.
\eea
 Eqs.~(\ref{axial}) and~(\ref{am_form}) therefore become
\bea
  \mat{p,s}{O_1^{\mu\nu\lambda}(0)}{\pp,s} & = &
   M a_0 \epsilon^{\mu\nu\lambda\beta}s_\beta,\nonumber \\
  \left. -i\derd{\Delta}{\nu}\mat{p,s}{O_2^{\mu\lambda}(0)}{\pp,s}\right|_{p=\pp}
   -(\nu \lra \lambda)
   & = & \frac{B_q(0)}{2M} p^{\{\mu}\epsilon^{\lambda\}\nu\beta\sigma}
   p_\beta s_\sigma -(\nu \lra \lambda)
  - M a_0 \epsilon^{\mu\nu\lambda\beta}s_\beta, \nonumber \\ \label{am_form2}\\
  \left. -i\derd{\Delta}{\nu}\mat{p,s}{O_3^{\mu\lambda}(0)}{\pp,s}\right|_{p=\pp}
   - (\nu \lra \lambda)
   & = & \frac{B_g(0)}{2M} p^{\{\mu}\epsilon^{\lambda\}\nu\beta\sigma}
   p_\beta s_\sigma -(\nu \lra \lambda).\nonumber
\eea
The decomposition in eqs.~(\ref{am_form2}) is one of the main points of this paper: There can be no
physically meaningful further decomposition of the proton spin
in terms of forward matrix elements, since we
only measure gauge-invariant quantities. Moreover, by adding the three equations together, one can see that the axial charge $a_0$ cancels from the sum-rule, so that only $B_{q/g}(0)$ appear in it. Thus, $a_0$ does not necessarily have to be interpreted as a measure of parton intrinsic spin. The best interpretation
that one can give for $B_{q/g}(0)$ are the total angular momentum of the
quarks and gluons $J_{q/g}$ respectively: 
\bea
  B_q(0) & = & J_q, \\
  B_g(0) & = & J_g. \nonumber
\eea
We refrain from decomposing $J_{q/g}$ further into spin plus orbital, since
it is not clear that it is meaningful to do so. We shall discuss this further
in section~\ref{sec_parton}.
The sum-rule
\be
 B_q(0) + B_g(0) = J_q + J_g = \frac{1}{2}
\ee
is assured by the non-renormalisation of the stress-energy tensor. This holds
in spite of the presence of the axial anomaly because in eq.~(\ref{am_form2})
it cancels out of the angular momentum sum-rule.

In the next section, we calculate the scaling of the operators
$O_{1\rightarrow 3}$ to one loop. Because this is done in the language of
operator vertices, as in the OPE analysis of DIS, it is important to show
how the above conclusions translate into the language of mixing of composite
operators. We shall argue that the axial anomaly affects not only the
renormalisation of $O_1$ but also of $xO_2$.
To do this, we appeal firstly
to some general properties of local gauge-invariant operators, and secondly
to the non-renormalisation of the AMC, which we have
verified to one loop in section~\ref{sec_leading}. We can thus write down
the general form of the mixing of the operators of the AMC.

First, we use the property that operators of the form $xO_i$ and $O_a$ when
inserted in forward matrix elements mix with a block triangular matrix:
\be
  \left( \begin{array}{c} O_a \\ xO_i \end{array} \right)_R
   = \left( \begin{array}{cc}  Z_{ab}^{-1} & 0 \\ Z_{ib}^{-1} & Z_{ij}^{-1}
   \end{array} \right)
   \left( \begin{array}{c} O_b \\ xO_j \end{array} \right)_B.
\label{x_mix}\ee
This is assured because local, gauge-invariant composite
operators with no factors of the co-ordinate $x^\mu$ only mix with other
such operators; the top right block of the matrix is therefore zero.
Second, we note that
because $O_3^{\mu\nu}$ is a symmetric tensor, it can only mix with the
symmetric operators $O_3^{\mu\nu}$ and $O_4^{\mu\nu}$. From eq.~(\ref{split}),
it then follows that $O_3^{\mu[\lambda} x^{\nu]}$ can only mix with
$O_3^{\mu[\lambda} x^{\nu]}$ and $O_1^{\mu\nu\lambda} +
O_2^{\mu[\lambda} x^{\nu]}$, when inserted in forward matrix elements.
Finally, given that the AMC is not renormalised,
the columns of the mixing matrix should add up to one. With these contraints,
the most general form of the matrix is as follows:
\be
  \left( \begin{array}{c} O_1^{\mu\nu\lambda} \\ O_2^{\mu[\lambda} x^{\nu]}
   \\ O_3^{\mu[\lambda} x^{\nu]} \end{array}\right)_B =
   \left(\begin{array}{ccc} 1 + X & 0 & 0 \\
                            Z - X & 1 + Z & -Y \\
                            -Z & -Z & 1 + Y \end{array}\right)
   \left(\begin{array}{c} O_1^{\mu\nu\lambda} \\ O_2^{\mu[\lambda} x^{\nu]}
   \\ O_3^{\mu[\lambda} x^{\nu]} \end{array}\right)_R,
\label{mix_ops}\ee
where $X = O(\alpha_s^2)$, $Y = O(\alpha_s)$ and $Z = O(\alpha_s)$. The piece
$X$ is due to the anomaly, which appears at two loops. It cancels between
$O_1$ and $xO_2$ in the AMC. In terms of the form factors, this mixing matrix
becomes
\be
  \left( \begin{array}{c} a_0 \\ B_q(0)
   \\ B_g(0) \end{array}\right)_B =
   \left(\begin{array}{ccc} 1 + X & 0 & 0 \\
                            0 & 1 + Z & -Y \\
                            0 & -Z & 1 + Y \end{array}\right)
   \left(\begin{array}{c} a_0 \\ B_q(0)
   \\ B_g(0) \end{array}\right)_R.
\label{mix_form}\ee

In the above, we have taken the AMC current not to be renormalised. This
we verify explicitly to one loop in the next section.


\section{One-Loop Evolution}
\label{sec_leading}

In this part, we present the mixing amongst the operators $O_1$, $O_2$, $O_3$
to leading order.
First, we discuss the technicalities of operator mixing.
The theory of mixing of gauge-invariant composite operators such as $O_1$ is
well understood. The extension to operators with a factor of $x^\mu$ is
straightforward. In particular, we show how operators of the form $xO_i$
can mix with operators of the form $O_a$.

Suppose we have
a local composite operator $B^{\mu\nu}(x)$ which mixes with a total
derivative operator $\partial_\alpha A^{\mu\nu\alpha}(x)$. The forward matrix
element of the total derivative operator would vanish. However, an operator
$x^\lambda B^{\mu\nu}(x)$ would correspondingly mix with
$x^\lambda \partial_\alpha A^{\mu\nu\alpha}(x)$
which is not a total derivative, but
which equals $\partial_\alpha (x^\lambda A^{\mu\nu\alpha}(x)) -
A^{\mu\nu\lambda}(x)$. The first term has vanishing forward
matrix element while the second is a local, composite
operator with no factor $x^\mu$ so gives an off-diagonal term in the mixing
matrix. To see this explicitly, we represent the forward matrix element
of $x^{[\nu} B^{\lambda]\mu}(x)$ as (See Appendix A.)
\be
  \lim_{\pp \rightarrow p} -i \derd{\Delta}{[\nu}
   \left[\raisebox{-0.5in}{\input{vert1.pstex_t}} \right],
\ee
where $\Gamma^{\mu\nu}$ is
the momentum-space vertex associated with the operator $B^{\mu\nu}$,
and the shaded blob is a four-point function, 1PI w.r.t.
the $p$ and $\pp$ legs.
As an example, consider the UV divergence of a loop
correction to $\Gamma^{\mu\nu}$:
\be
  -i \derd{\Delta}{[\nu} \left[ \raisebox{-0.5in}{\input{vert2.pstex_t}}
   \right] \stackrel{{\rm large} \phantom{a}l^2}{\longrightarrow}
  -i \derd{\Delta}{[\nu} \left[ \raisebox{-0.5in}{\input{vert3.pstex_t}} \right].
\ee
The divergence piece takes the form, in dimensional regularisation, of
$1/\epsilon$ pole pieces:
\be
  \Gamma^{\mu\nu}(k,\Delta,\epsilon) = \Gamma^{\mu\nu}_0(k,\epsilon) +
   i\Delta_\alpha \Gamma^{\mu\nu\alpha}_1(k,\epsilon) + O(\Delta^2).
\ee
The $O(\Delta)$ pieces, $\Delta_\alpha \Gamma^{\mu\nu\alpha}_1$,
correspond to mixing with a total derivative operator
$\partial_\alpha A^{\mu\nu\alpha}$.
The $\Delta$ differentiation acts on the blob, and then on the pole
pieces. The first corresponds to mixing with another operator of the form
$xO_i$, while the second gives
\be
  -i \derd{\Delta}{[\nu} \Gamma^{\lambda]\mu}(k,\Delta,\epsilon) =
   \Gamma_1^{\lambda\mu\nu}(k,\epsilon)
   - (\nu \lra \lambda) + O(\Delta).
\ee
Thus, when considering the forward matrix elements, the mixing
of an operator $B^{\mu\nu}$ with a total derivative operator
$\partial_\alpha A^{\mu\nu\alpha}$ corresponds to mixing of
$B^{\mu[\lambda} x^{\nu]}$ with $A^{\mu[\lambda \nu]}$.

Now we give the one-loop mixing of the operators in the gauge-invariant
decomposition of the AMC. The basis of operators required is the following:
\bea
  O_1^{\mu\nu\lambda} & = & \frac{1}{2} \epsilon^{\mu\nu\lambda\alpha}
   \psibar \gamma_\alpha \gamma^5 \psi, \nonumber \\
  O_2^{\mu\nu} & = & \psibar\gamma^\mu i\hD{\nu} \psi, \nonumber \\
  O_3^{\mu\nu} & = & F^{\mu\alpha}F_\alpha^{\phantom{a}\nu}, \nonumber \\
  O_4^{\mu\nu} & = & \frac{1}{2} \psibar\gamma^{\{\mu} i\hD{\nu\}} \psi.
\eea
These operators also mix with $g^{\mu\nu}O_5$ and $g^{\mu\nu}E_1$, where
\bea
  O_5 & = & F^{\alpha\beta}F_{\alpha\beta}, \nonumber \\
  E_1 & = & i\psibar\shD\psi,
\eea
but we shall not consider these since
not only are they projected out in the definition
of $J^{ij}$ in eq.~(\ref{spin_vector}),
but also the first derivatives of their matrix elements vanish in the
forward limit. We do not require operators of lower
dimension, since we are neglecting quark masses.
Note that we require not just $O_{1\rightarrow 3}$ but also $O_4$, which is
$O_2$ with its indices symmetrised. The fact that $O_2$ and $O_3$ will mix
with $O_4$ does not spoil the non-renormalisation of the g.i. decomposition
because of eq.~(\ref{split}).

Inserting the operators in the Green functions with two gluon or two quark
legs, we calculate the pole pieces to one loop. Details are to be
found in Appendix B. We find that the matrix $Z_{ij}$ required to renormalise
the bare operators is
\bea
  (O_i)_B & = & Z_{ij} (O_j)_R,\nonumber \\
  \left( \begin{array}{c} \partial_\lambda O_1^{\mu\nu\lambda} \\
   O_2^{\mu\nu} \\ O_3^{\mu\nu} \\ O_4^{\mu\nu} \end{array} \right)_B
   & = & \left(
   \begin{array}{cccc} 1 & 0 & 0 & 0 \\
         0 & 1 & \frac{2}{3}n_f\frac{\alpha_s}{4\pi}\frac{1}{\epsilon}
           & -\frac{8}{3}C_F\frac{\alpha_s}{4\pi}\frac{1}{\epsilon} \\
         0 & 0 & 1 - \frac{2}{3}n_f\frac{\alpha_s}{4\pi}\frac{1}{\epsilon}
           & \frac{8}{3}C_F\frac{\alpha_s}{4\pi}\frac{1}{\epsilon} \\
         0 & 0 & \frac{2}{3}n_f\frac{\alpha_s}{4\pi}\frac{1}{\epsilon}
           & 1 - \frac{8}{3}C_F\frac{\alpha_s}{4\pi}\frac{1}{\epsilon}
   \end{array} \right)
   \left( \begin{array}{c}\partial_\lambda O_1^{\mu\nu\lambda} \\
   O_2^{\mu\nu} \\ O_3^{\mu\nu} \\ O_4^{\mu\nu} \end{array} \right)_R.
\label{mix_op}\eea
The angular momentum current is not renormalised to this order. Note also that
the counterterms for $O_2$ are the same as those for $O_4$. Thus, the
renormalisation of $xO_2$ and $xO_3$ follows that for the stress-energy tensor
$T^{\mu\nu}$.
At the two-loop level, we anticipate that this behaviour will be modified
by the axial anomaly, where $O_1$ mixes with itself only~\cite{ET}:
\be
  (O_1^{\mu\nu\lambda})_B = Z_{11}(O_1^{\mu\nu\lambda})_R.
\ee
If the AM current is not to be renormalised, then $xO_2$ will have to mix
with $O_1$.

Using eq.~(\ref{split}), when the operators are inserted in forward
matrix elements, eq.~(\ref{mix_op}) becomes
\bea
  \left( \begin{array}{c} O_1^{\mu\nu\lambda} \\
   O_2^{\mu[\nu}x^{\lambda]} \\ O_3^{\mu[\nu}x^{\lambda]} \end{array} \right)_B
   & = & \left(
   \begin{array}{ccc} 1 & 0 & 0\\
         - \frac{8}{3}C_F\frac{\alpha_s}{4\pi}\frac{1}{\epsilon}
           & 1 - \frac{8}{3}C_F\frac{\alpha_s}{4\pi}\frac{1}{\epsilon}
           & \frac{2}{3}n_f\frac{\alpha_s}{4\pi}\frac{1}{\epsilon}\\
         \frac{8}{3}C_F\frac{\alpha_s}{4\pi}\frac{1}{\epsilon}
           & \frac{8}{3}C_F\frac{\alpha_s}{4\pi}\frac{1}{\epsilon}
           & 1 - \frac{2}{3}n_f\frac{\alpha_s}{4\pi}\frac{1}{\epsilon}
   \end{array} \right)
   \left( \begin{array}{c} O_1^{\mu\nu\lambda} \\
   O_2^{\mu[\nu}x^{\lambda]} \\ O_3^{\mu[\nu} x^{\lambda]} \end{array} \right)_R.
\label{mix_op_mod}\eea

If we now take the forward matrix elements of $O_1$, $xO_2$ and $xO_3$ as defined by
eq.~(\ref{for_mat}) we obtain,
to $O(\alpha_s)$,
\bea
  \frac{d}{dt} a_0 & = & 0 \nonumber \\
  \frac{d}{dt} B_q & = & \frac{\alpha_s}{4\pi} \left[
   -\frac{8}{3}C_F B_q
   + \frac{2}{3} n_f B_g \right]
   \nonumber \\
  \frac{d}{dt} B_g & = & \frac{\alpha_s}{4\pi} \left[
   \frac{8}{3}C_F B_q
   - \frac{2}{3} n_f B_g \right].
\label{mix_dist_worka}\eea


\section{The Parton Model Decomposition}
\label{sec_parton}

Since the angular momentum sum rule only involves two form factors and is
independent of the axial charge and the anomaly associated with it, the
best decomposition of the proton spin that we can have is
\be J_q + J_g = \frac{1}{2}. \ee
In the context of a parton model, we would like to be able to write
\be \frac{1}{2} \Delta q + \Delta g + L_q + L_g = \frac{1}{2}, \ee
so that
\bea
  J_q = \frac{1}{2} \Delta q + L_q, \nonumber \\
  J_g = \Delta g + L_g.
\eea
This is not a natural thing to do since we would like
to identify $\Delta q$ with the axial charge, which multiplies a different
Lorentz structure from $J_{q/g}$ in the form factor decomposition. In other
words, such a decomposition is dependent on the Lorentz frame.

Nevertheless, to make contact with the results of HJT~\cite{HJT}, we also relate the three form factors in terms of parton model quantities.
As remarked above, the central problem is to relate
four observables, $\Delta q$, $\Delta g$, $L_q$ and $L_g$ to just three
form factors, not least because we can imagine each scaling in a distinct way.
We will argue, however, that because it is possible to define
$\Delta q$ to be scale-independent, then we may determine the scaling of
$\Delta g$, $L_q$ and $L_g$ gauge-invariantly in terms of the scaling of the
three gauge-invariant operators in eq.~(\ref{gi_decomp}).
The problem is the same as that in polarised DIS, where there are
two observables, $\Delta q$ and $\Delta g$ to relate to just one form factor,
$a^0$. Below, we briefly 
review polarised DIS and the role of the axial anomaly, and how $\Delta q$
can be constrained to be scale-independent in a certain class of
renormalisation schemes. We then identify form factors with parton model
quantities, and rederive the result of HJT~\cite{HJT} for the mixing of
$\Delta q$, $\Delta g$, $L_q$ and $L_g$. Thus we will have shown that an
operator definition of these quantities does in fact lead to the HJT
mixing.

Let us begin with the first moment $\Gamma_1$ of the polarised structure
function $g_1(x,Q^2)$ as measured in
polarised DIS. For large $Q^2$, $\Gamma_1$ factorises in the following way:
\bea
  \Gamma_1 = \int_0^1 dx \, g_1(x,Q^2)
   & = & \frac{\langle e^2\rangle}{2}
   \left[ C_{\rm NS}^1(\alpha_s) \Delta q_{\rm NS}
   + C_{\rm S}^1(\alpha_s) \Delta q + 2n_f C_g^1(\alpha_s) \Delta g \right],
   \nonumber \\
  \Delta q_{\rm NS} & = & \frac{1}{n_f} \sum_{i=1}^{n_f}
   \left(\frac{e^2_i}{\langle e^2\rangle} - 1\right) ( \Delta q_i +
   \Delta\overline{q}_i ), \label{fact}\\
  \Delta q & = & \sum_{i=1}^{n_f} ( \Delta q_i +
   \Delta\overline{q}_i ), \nonumber
\eea
where $\langle e^2 \rangle = \frac{1}{n_f} \sum_i e^2_i$,
$\Delta q_i$, $\Delta \overline{q}_i$ and $\Delta g$ are the first
moments of the polarised
quark, antiquark and gluon parton distribution functions
(PDFs) respectively, and
$C^1(\alpha_s)$ are the first moments of the co-efficient functions.
Both the co-efficient functions and the PDFs are scheme dependent.
At leading order, parton amplitude calculations show that the scaling of
$\Delta q$ and $\Delta g$ are given by
\be
  \frac{d}{dt}\left(\begin{array}{c} \Delta q \\ \Delta g \end{array}\right)
   = \frac{\alpha_s}{4\pi} \left(\begin{array}{cc} 0 & 0 \\ 3C_F & \beta_0
   \end{array}\right)
   \left(\begin{array}{c} \Delta q \\ \Delta g \end{array}\right),
\ee
where $C_F = \frac{4}{3}$ and $\beta_0 = 11 - \frac{2}{3}n_f$.
The eigenvectors of this evolution are $\Delta q$, which does not evolve,
and $\Delta q - 2 n_f \frac{\alpha_s}{4\pi} \Delta g$. It then
follows that there should exist a class of renormalisation schemes, the Adler-Bardeen (AB) schemes~\cite{BFR}, where
these two eigenvectors are the same to all orders. In such a scheme, it is the
second of the eigenvectors which may then appear in the factorisation
in eq.~(\ref{fact}): By contrast, in the
$\MS$ scheme, the gluon co-efficient $C_g^1(\alpha_s)$ vanishes to
$O(\alpha_s^2)$, so that the only flavour singlet piece in $\Gamma_1$ is
the contribution from $\Delta q$. The factorisation
can also be expressed in terms of the operator product expansion (OPE). Since
the only flavour singlet operator with the right dimension is the axial
current,
\be
  \mat{p,s}{O^{\mu\nu\lambda}_1}{p,s} =
   M\epsilon^{\mu\nu\lambda\rho} s_\rho a_0(t),
\ee
we have, for the $\MS$ scheme, $a^0(t) = \Delta q(t)$. That is,
$\Delta g$ does not contribute, while $\Delta q(t)$ depends on the scale
$t$ because of
the axial anomaly. However, in the class of AB schemes~\cite{BFR},
the gluon co-efficient is $C_g^1(\alpha_s) = -\frac{\alpha_s}{4\pi}
+ O(\alpha^2_s)$, while $C_{\rm NS}^1(\alpha_s) = C_{\rm S}^1(\alpha_s)
= 1 - \frac{\alpha_s}{\pi} + O(\alpha^2_s)$, and we may identify~\cite{AR}
\be
  a_0(t) = \Delta q - 2n_f \frac{\alpha_s}{4\pi} \Delta g(t).
\label{AR_decomp}\ee
Thus, the parton model interpretation of the axial current is that it measures both quark spin and gluon spin.
The existence of schemes where this identification holds to all orders
is assured by the Adler-Bardeen theorem~\cite{AB}. $\Delta q$ is
scale-independent to all orders, while the scaling of $\Delta g(t)$ is
determined by the scaling of the axial current.

Now we make the parton model identifications according to the AB scheme
interpretation:
\bea
  a_0 & = & \Delta q - 2n_f \frac{\alpha_s}{4\pi} \Delta g, \nonumber \\
  B_q(0) & = & \frac{1}{2} \Delta q + L_q, \label{observ}\\
  B_g(0) & = & \Delta g + L_g. \nonumber
\eea
The one-loop mixing derived in (\ref{mix_dist_worka}) then gives us
\bea
  \frac{d}{dt} \Delta q & = & 0 \nonumber \\
  \frac{d}{dt} L_q(t) & = & \frac{\alpha_s}{4\pi} \left[
   -\frac{8}{3}C_F(\frac{1}{2}\Delta q + L_q)
   + \frac{2}{3} n_f (\Delta g + L_g) \right]
   \nonumber \\
  \frac{d}{dt} (L_g(t) +
   \Delta g(t)) & = & \frac{\alpha_s}{4\pi} \left[
   \frac{8}{3}C_F(\frac{1}{2}\Delta q + L_q)
   - \frac{2}{3} n_f (\Delta g + L_g) \right]
\label{mix_dist_workb}\eea

Thus $d\Delta g/dt$ remains undetermined. For this we require the two-loop
evolution of $a^0$~\cite{Kod},
\be
  \frac{d}{dt} a^0(t) = \gamma a^0(t), \phantom{********}
   \gamma = - n_f \frac{\alpha_s^2}{2\pi^2}.
\ee
With the identification~(\ref{AR_decomp}), this implies
\be
  \frac{d}{dt} \Delta g(t) = \frac{\alpha_s}{4\pi} \left[
   3C_F \Delta q + \beta_0 \Delta g \right].
\label{mix_dist_workc}\ee
Combining~(\ref{mix_dist_workb})
and~(\ref{mix_dist_workc}), we have the evolution equation, in agreement with
ref.~\cite{HJT}:
\be
  \frac{d}{dt} \left(\begin{array}{c} \Delta q \\ \Delta g \\ L_q \\ L_g
   \end{array}\right) = \frac{\alpha_s}{4\pi} \left(
   \begin{array}{cccc} 0 & 0 & 0 & 0 \\
                3C_F & \beta_0 & 0 & 0 \\
          -\frac{4}{3}C_F & \frac{2}{3}n_f & -\frac{8}{3}C_F & \frac{2}{3}n_f\\
          -\frac{5}{3}C_F & -11 & \frac{8}{3}C_F & -\frac{2}{3}n_f
   \end{array}\right)\left(\begin{array}{c}\Delta q \\ \Delta g \\ L_q \\ L_g
   \end{array}\right),
\ee
with
\be
  \frac{d}{dt}\left( \frac{1}{2}\Delta q + \Delta g + L_q + L_g \right) = 0.
\ee
The general solution to these coupled equations is obtained in ref.~\cite{HJT},
whence in the asymptotic limit $t \rightarrow \infty$ the partitioning
of total quark and gluon angular momentum is $\frac{1}{2} \Delta q + L_q :
\Delta g + L_g = 3n_f : 16$. That result is a direct consequence
of how the form
factors $B_{q/g}(0)$ mix only with each other. This is the same partitioning
between quarks and gluons of the first moments of the unpolarised PDFs.


\section{Measurement of Off-Forward Matrix Elements}
\label{sec_meas}

As was remarked above, of the three form-factors of operators appearing in the angular momentum sum-rule, only one, the axial charge, can be measured in polarised
DIS. To measure the other two, one requires off-forward matrix elements of
$O^{\mu\nu}_3$ and $O^{\mu\nu}_4$. As has been proposed in
ref.~\cite{Ji,Ji2}, these can be extracted from a class of hard, exclusive
processes, generically $\gamma^\ast p \rightarrow X p$. For large $Q^2$, these
factorise into perturbatively calculable pieces, skewed PDFs and other
non-perturbative pieces depending on the choice of the final-state species
$X$. The skewed PDFs can be represented as off-forward proton matrix elements
of a time-ordered bilocal operator product. The first moments of these are
then equal to off-forward matrix elements of local composite operators.
Factorisation theorems have been treated for Deeply Virtual Compton
Scattering (DVCS)~\cite{Rad,OJ,CFS},
$\gamma^\ast p \rightarrow \gamma p$, where
the final-state photon is real, and also for meson production
$\gamma^\ast p \rightarrow p V$.
For completeness, we review this work briefly, pointing out that we can
measure proton spin components in the combinations $B_q(0) = J_q$ and
$B_g(0) = J_g$~\cite{Ji,Ji2}. This supplements the isolation of $\Delta q$
and $\Delta g$ in polarised DIS and of $\Delta g$ in open charm production.
\begin{figure}
\centering{\input{exclusive.pstex_t}}
\caption{Definitions of four-momenta for hard, exclusive processes.}
\label{fig_excl}\end{figure}
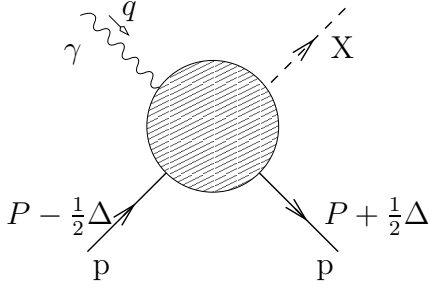

Consider then the process $\gamma^\ast p \rightarrow X p$, where we leave
the species $X$ unspecified. The amplitude $\mathcal{M}_X$ for
the process factorises as follows:
\be
  \mathcal{M}_X(q, P, \Delta) \stackrel{Q^2 \rightarrow \infty}{\longrightarrow}
   \sum_i \sum_\sigma \int du \, C^i_\sigma (Q^2, \frac{x}{u}, \xi)
   f_{i/{\rm p}}(u, \xi, \Delta^2) D_{{\rm X}/\sigma}.
\ee
The four-momenta are defined as in fig.~\ref{fig_excl}. The sum over $i$ is
over partons originating in the proton p; that over $\sigma$ is for, say,
degrees of freedom going into the species X. The Lorentz-invariant kinematic
variables are
\bea
  Q^2 & = & -q^2, \\
  x & = & \frac{Q^2}{2q\cdot P}, \\
  \xi & = & \frac{q\cdot\Delta}{2 q\cdot P}
   \stackrel{Q^2 \rightarrow \infty}{\longrightarrow} \frac{\Delta^+}{2P^+}.
\eea
The unpolarised skewed PDFs, which we only need to consider, are given by
\bea
  f_{\rm q/p}(u, \xi, \Delta) & = & \int \meas{y^-}{} e^{-i(u+\xi/2)P^+y^-}
   \mat{P+\frac{1}{2}\Delta}{T(\psibar(y^-)\gamma^+\mathcal{P}(y^-,0)\psi(0))}%
   {P-\frac{1}{2}\Delta}, \nonumber \\
  P^+ \; f_{\rm g/p}(u, \xi, \Delta) & = & \int \meas{y^-}{} e^{-i(u+\xi/2)P^+y^-}
   \mat{P+\frac{1}{2}\Delta}{T(F^{+\alpha}(y^-)%
   \mathcal{P}(y^-,0)F_\alpha^{\phantom{a}+}(0))}
   {P-\frac{1}{2}\Delta}, \nonumber \\
\eea
where $\mathcal{P}(x,y)$ is the path-ordered string operator which renders
a bilocal operator depending on two points $x$ and $y$ gauge-invariant:
\be
  \mathcal{P}(x,y) = P \exp \left(-ig \int_x^y ds\cdot A \right).
\ee
In the forward limit, $\Delta \rightarrow 0$, these tend toward the usual PDFs
with the correspondence
\bea
  f_{\rm q/p}(u, \xi, \Delta) \stackrel{\Delta \rightarrow 0}{\longrightarrow}
   f_{\rm q/p}(u), \nonumber \\
  f_{\rm g/p}(u, \xi, \Delta) \stackrel{\Delta \rightarrow 0}{\longrightarrow}
   u\, f_{\rm g/p}(u).
\eea

The skewed PDFs have support for $-1 < u < 1$, so their first moments w.r.t.
$u$ are
\bea
  (P^+)^2\; \int_{-1}^1 du\, u\, f_{\rm q/p}(u, \xi, \Delta) & = &
   \mat{P+\frac{1}{2}\Delta}{\psibar\gamma^+ i\hD{+} \psi(0)}%
   {P-\frac{1}{2}\Delta}, \nonumber \\
  (P^+)^2\; \int_{-1}^1 du\, f_{\rm q/p}(u, \xi, \Delta) & = &
   \mat{P+\frac{1}{2}\Delta}{F^{+\alpha}%
   F_\alpha^{\phantom{a}+}(0)}{P-\frac{1}{2}\Delta}.
\eea
Decomposing the matrix elements as in section~\ref{sec_role} we can relate the
form factors at zero momentum transfer $B_q(0)$ and $B_g(0)$ to the first
derivatives of the first moments of the PDFs:
\bea
  \left. -iP^+\derd{\Delta}{\mu} \int_{-1}^1 du\, u\, f_{\rm q/p}(u, \xi, \Delta)
   \right|_{\Delta = 0} & = & \frac{B_q(0)}{M} \epsilon^{+\mu\alpha\beta}
   P_\alpha s_\beta, \\
  \left. -iP^+\derd{\Delta}{\mu} \int_{-1}^1 du\, f_{\rm g/p}(u, \xi, \Delta)
   \right|_{\Delta = 0} & = & \frac{B_g(0)}{M} \epsilon^{+\mu\alpha\beta}
   P_\alpha s_\beta,
\eea
where
\bea
   B_q(0) & = & J_q, \\
   B_g(0) & = & J_g.
\eea
Thus, if the spin vector $s^\mu$-dependent part of the first derivative
of the first moments of the skewed PDFs can be extracted from experiment, then
this gives information on the total quark and total gluon angular momentum.


\section{Gauge-Non-Invariant Decompositions}
\label{sec_gninv}

Here we discuss further the gauge-non-invariant decomposition of the AMC
of eq.~(\ref{nongi_decomp}), in particular the possible merits
of relating this to the four components $\Delta q$, $\Delta g$,
$L_q$ and $L_g$ in one particular choice of gauge and Lorentz frame. The
natural choice for a parton model interpretation would be the axial gauge
$A^+ = 0$ and the infinite momentum frame, that is, the rest frame of the
virtual photon in DIS in the limit $Q^2 \rightarrow \infty$. Indeed, one
finds attempts~\cite{BJ} to relate moments of angular momentum PDFs,
$\Delta q(x)$, $\Delta g(x)$, $L_q(x)$ and $L_g(x)$ to local, composite
operators in the gauge $A^+ = 0$. However, one is henceforth restricted
to a particular choice of gauge and quantisation procedure, viz. light-cone
quantisation, which is inconvenient beyond low orders in perturbation theory.
We also discuss attempts to relate $\Delta g$ to the forward matrix elements
of the Chern-Simons current $k^\mu$, and review the problem which befalls this
identification, namely the coupling of $k^\mu$ to unphysical pseudoscalar
poles.

To begin, suppose that we quantise QCD in the infinite momentum frame with the
gauge $A^+ = 0$, i.e. use light-cone quantisation. In this picture, the only
propagating degrees of freedom are the transverse gluon polarisations,
$A^i_\perp$, and we do not require any gauge-fixing fields or ghosts. Since the
fields are quantised with equal $x^+$
commutation relations,\footnote{%
Our convention is that $x^\pm = \frac{1}{\sqrt{2}}(x^0 \pm x^3)$.}
then the canonical rotation generator is
\be
  J^{ij}_{(+)} = \int dx^- \, d^2x^\perp \, M^{+ij},
\ee
with
\be
  M^{\mu\nu\lambda} = \frac{1}{2}\epsilon^{\mu\nu\lambda\alpha}
   \psibar \gamma_\alpha \gamma^5 \psi
   + i\psibar \gamma^\mu x^{[\nu}\partial^{\lambda]} \psi
   - F^{\mu[\nu}A^{\lambda]} -
   F^{\mu\alpha}x^{[\nu}\partial^{\lambda]}A_\alpha.
\ee
Thus,
\bea
  J^{ij}_{(+)} & = & \int dx^- \, d^2x^\perp \, \left[\sqrt{2}\psi^\dagger_+
   \sigma^{ij}\psi_+ + i\sqrt{2}\psi^\dagger_+x^{[i}\partial^{j]}\psi_+ \right.
   \nonumber \\
  & - & \left. \partial^+ A^{[i} A^{j]}
   + \partial^+ A^k_\perp x^{[i}\partial^{j]}A^k_\perp \right],
\label{lc_decomp}\eea
where $\psi_+ = \Lambda^+ \psi = \frac{1}{\sqrt{2}}\gamma^0\gamma^+\psi$.
This equation is of the same form as eq.~(\ref{canon}). The difficulty
here now is that the four operators are not even invariant under the residual
gauge transformations which leave the condition $A^+ = 0$ unchanged.
In ref.~\cite{BJ}, however, definitions of $\Delta q(x)$, $\Delta g(x)$,
$L_q(x)$ and $L_g(x)$ are constructed in the axial gauge which are invariant
under residual gauge transformations. The first moments of these,
$\int_{-1}^1 \Delta q(x) dx$ etc., correspond to the four operators in
eq.~(\ref{lc_decomp}), but with $x^{[i}\partial^{j]}$ replaced by
$x^{[i}\mathcal{D}^{j]}$, with $\mathcal{D}^\mu$ being the derivative
covariant under the residual gauge transformation. It is not altogether clear
that those modified operators add up to give a conserved current. Moreover,
one is restricted to a particular gauge, so if one encounters technical
difficulties in perturbative calculations of the different contributions
to the proton spin,
as reported in ref.~\cite{HJT2}, one has no room for manoeuvre.

The other feature of axial gauges is that the $(+12)$ component of
$F^{\mu[\nu}A^{\lambda]}$ is related to the Chern-Simons current $k^\mu$.
\be
  k^\mu = \frac{\alpha_s}{2\pi} \epsilon^{\mu\nu\rho\sigma} {\rm tr} \left[
   A_\nu \partial_\rho A_\sigma - \frac{2ig}{3} A_\nu A_\rho A_\sigma\right].
\ee
Indeed, putting $A^+ = 0$, we have
$k^+ = \frac{\alpha_s}{4\pi} \partial^+ A^{[1}A^{2]}$. Note that this is a
frame-dependent and gauge-dependent statement. Some
authors~\cite{AR,CCM,Bass} have
tried to identify the forward matrix element of $k^\mu$ with
$-\frac{\alpha_s}{4\pi} \Delta g$ in such a way that $\Delta g$ is at least
a Lorentz scalar form factor: From the anomalous Ward identity,
which is not renormalised,
\be
  \partial_\mu j^\mu_5 - 2n_f Q \simeq 0,
\ee
with
$Q = \frac{\alpha_s}{8\pi} {\rm tr}F^{\mu\nu}F_{\mu\nu} = \partial_\mu k^\mu$
and $j^\mu_5 = \psibar\gamma^\mu\gamma_5\psi$, one deduces that
$j^\mu_5 - 2n_f k^\mu$ is not renormalised, so that with
\bea
  \mat{p,s}{k^\mu}{\pp,s} & = & 2Ms^\mu a^0_g(\Delta^2)
  + s\cdot\Delta\Delta^\mu p_g(\Delta^2),
   \nonumber \\
  \mat{p,s}{(j^\mu_5 - 2n_f k^\mu)}{\pp,s} & = & 2Ms^\mu a^0_q(\Delta^2)
   + s\cdot\Delta\Delta^\mu p_q(\Delta^2),
\eea
we have
\be
  \mat{p,s}{j^\mu_5}{p,s} = 2Ms^\mu (a^0_q(0) + 2n_fa^0_g(0)),
\ee
so that
\bea
  \Delta q & = & a^0_q(0) \nonumber \\
  -\frac{\alpha_s}{4\pi} \Delta g & = & a^0_g(0).
\eea
This is true provided, as should be the case, that
$\lim_{\Delta \rightarrow 0}[2n_f p_g(\Delta^2) + p_q(\Delta^2)]$ is finite.

However, in covariant gauges, the presence of an unphysical
massless pseudoscalar pole~\cite{Kugo} coupling to $k^\mu$ means that as
$\Delta \rightarrow 0$
\bea
  2n_f p_g(\Delta^2) & \rightarrow & \frac{1}{\Delta^2}, \nonumber \\
  p_q(\Delta^2) & \rightarrow & -\frac{1}{\Delta^2}.
\eea
In other words, the forward matrix element of the Chern-Simons current is
singular, when treated non-perturbatively.
In axial gauges, $n\cdot A = 0$, this pole becomes a $\frac{1}{n\cdot\Delta}$
pole, so that the Lorentz structure $n^\mu/(n\cdot\Delta)$ is present instead
of $\Delta^\mu/(\Delta^2)$ in the limit $\Delta \rightarrow 0$. This has
been shown for the Schwinger model explicitly in ref.~\cite{Man} and for
QCD in ref.~\cite{BB}. That said, for the gauge $A^+ = 0$, $n^+ = 0$, so no
such pole is present in $\mat{p,s}{k^+}{p,s}$; only for one choice of gauge and
frame can the forward matrix element be well defined.

In conclusion, defining observables in terms of matrix elements of gauge-non-invariant operators requires reference to a specific gauge, which is problematic. The matrix elements are also singular in the forward limit. Since the gauge-non-invariant approach is sufficient for our purposes, there is no need for further decomposition of the AMC.


\section{Discussion and Conclusions}

In summary, we have decomposed the conserved angular momentum current keeping it gauge invariant as far as possible. We have then decomposed the forward matrix elements of these operators into scalar form factors, and shown that the axial charge drops out of the angular momentum sum-rule. This is important because it provides evidence that the axial charge is not straightforwardly interpreted as a measure of parton intrinsic spin. Indeed, the parton model interpretation which we have discussed in section~\ref{sec_parton}
is such that $\Delta q$ and $\Delta g$ appear in form factors multiplying two different Lorentz structures, one of which cancels out of the sum-rule.
One should thus take the view that although it is meaningful to split angular
momentum into quark and gluon pieces, a further split into spin and orbital
pieces is frame dependent and therefore not a property of the proton itself.

We have derived the mixing of the three gauge-invariant operators in the AMC,
and calculated this explicitly to one loop. In general, we have shown how the
axial anomaly cannot affect the renormalisation of the AMC, and this
is reflected at the level of form factors by the decoupling of the axial
charge from the angular momentum sum rule.

Despite having argued that it is unnatural to do so, we have also derived the
mixing of $\Delta q$, $\Delta g$, $L_q$
and $L_g$ by relating these to form factors of the gauge-invariant
composite operators, and found
results in agreement with what is obtained by taking the first moments of
splitting functions. The scaling
of $\Delta g$ has been determined from the scaling of the axial charge provided $\Delta q$ is constrained to be scale-independent.
This has enabled us to manage with just three operators. Thus, the
mixing matrix of HJT can be derived in an explicitly gauge-invariant way in
terms of local, composite operators. This means that we should not need
to consider using gauge-non-invariant operators.


\section*{Appendix A}

In this appendix we derive an expression for the forward matrix element of an
operator containing a factor $x^\mu$. This must be done with care otherwise
we end up with a derivative of the momentum-conserving delta-function.
To see this, consider the off-forward nucleon matrix element of an operator
$x^{[i} B^{j]0}$:
\bea
  \int \dx \mat{p}{x^{[i} B^{j]0}(x)}{\pp} & = &
   \int \dx e^{i(\pp - p)\cdot x} x^{[i} \mat{p}{B^{j]0}(0)}{\pp} \nonumber \\
  & = & i(2\pi)^3  \derd{\Delta}{[i} \left( \delta^3(p-\pp) \right)
   \mat{p}{B^{j]0}(0)}{\pp},
\label{ill_def}\eea
where $\mat{p}{B(0)}{\pp}$ is a Green function amputated w.r.t. its $p$ and
$\pp$ legs and where $P$ is held constant in the derivative.
We may fix this by replacing $\ket{\pp}$ with a wavepacket,
\be
  \ket{\Phi} = \int \meas{\Delta}{3} \Phi(\pp) \ket{\pp},
\ee
such that $\Phi(\pp)$ has no azimuthal dependence about the k-direction,
perpendicular to the i- and j-directions. Then,
integrating by parts, we have
\bea
  \int \dx \mat{p}{x^{[i} B^{j]0}(x)}{\Phi}
   & = & \left. -i\derd{\Delta}{[i}(\Phi(\pp))\right|_{\pp=p}
   \mat{p}{B^{j]0}(0)}{p}\nonumber \\
  & - & \Phi(p) \left. i\derd{\Delta}{[i} \mat{p}{B^{j]0}(0)}{\pp}\right|_{\pp=p}.
\eea
The first term on the r.h.s. vanishes since $\mat{p}{B^{0j}(0)}{p} =
p^0p^j \mat{p}{\widetilde{B}}{p}$, with $\mat{p}{\widetilde{B}}{p}$ the reduced
matrix element, so that we get a factor $p^{[j}\derd{p}{i]}\Phi(p)$ which
vanishes by construction --- i.e. the wavepacket carries no OAM.
Thus we consider only the term
\be
  \Phi^{-1}(p) \int \dx \mat{p}{x^{[i}B^{j]0}(x)}{\Phi}
   = -i \derd{\Delta}{[i} \left. \mat{p}{B^{j]0}(0)}{\pp} \right|_{\pp=p}.
\label{for_def}\ee
Notice that the l.h.s. reduces to $\mat{p}{B(0)}{p}$ if we remove the factor
$x^\mu$ and talk about the matrix element of an ordinary, local, composite
operator.

Next, we show that matrix elements of operators such as
$\partial^\alpha(x^\beta B(x))$ vanish, so that the forward matrix element
of the total derivative term in eq.~(\ref{gi_sep}) can be neglected.
We get
\bea
  \int \dx \mat{p}{\partial^\alpha(x^\beta B(x))}{\Phi}
  & = & -(2\pi)^3 \int \meas{\Delta}{3}(p-\pp)^\alpha\delta^3(p-\pp) \left[
   \derd{\Delta}{\beta} (\Phi(\pp)) \mat{p}{B(0)}{\pp}\right. \nonumber \\
  & + & \left. \Phi(\pp) \derd{\Delta}{\beta} \mat{p}{B(0)}{\pp}\right],
\eea
which is zero.


\section*{Appendix B}

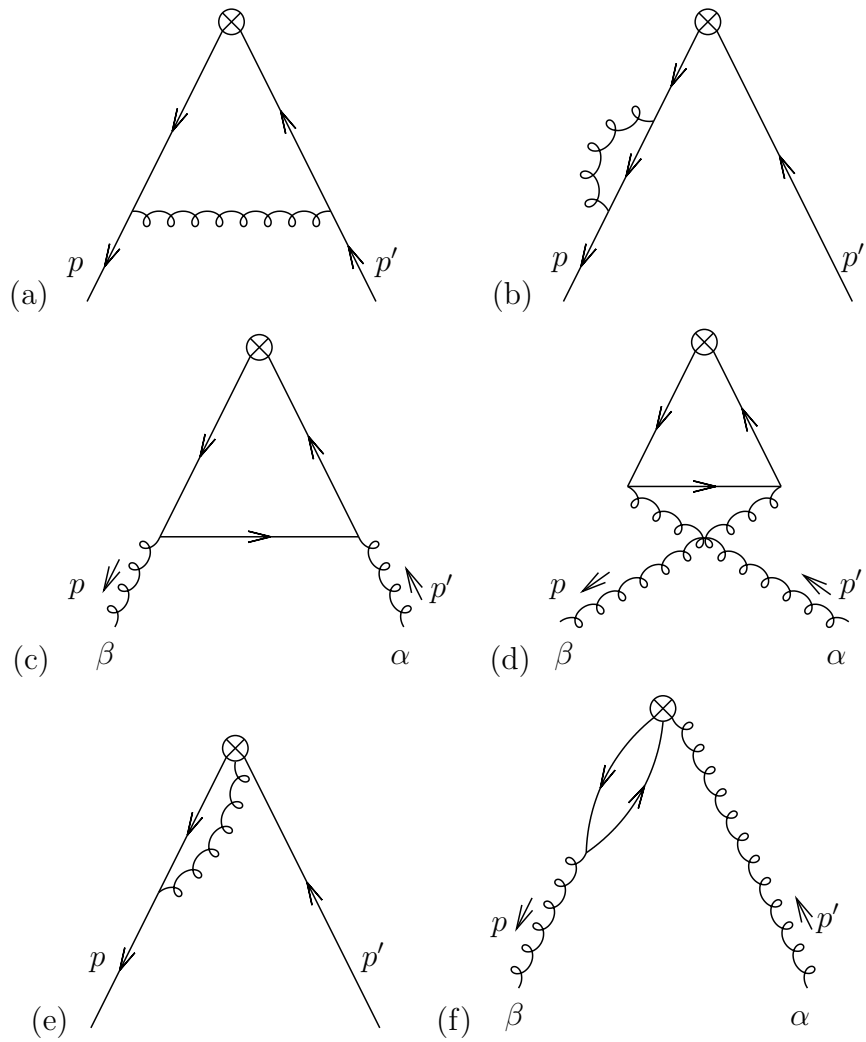
\begin{figure}
\leavevmode
\begin{center}
  (a)\hspace{0.1in}\input{diaga.pstex_t}\hspace{0.5in}
  (b)\hspace{0.1in}\input{diagb.pstex_t} \\ \vspace{0.1in}
  (c)\hspace{0.1in}\input{diagc.pstex_t}\hspace{0.2in}
  (d)\hspace{0.1in}\input{diagd.pstex_t} \\ \vspace{0.1in}
  (e)\hspace{0.1in}\input{diage.pstex_t}\hspace{0.2in}
  (f)\hspace{0.1in}\input{diagf.pstex_t}
\caption{One-loop diagrams}
\label{diags1}
\end{center}
\end{figure}

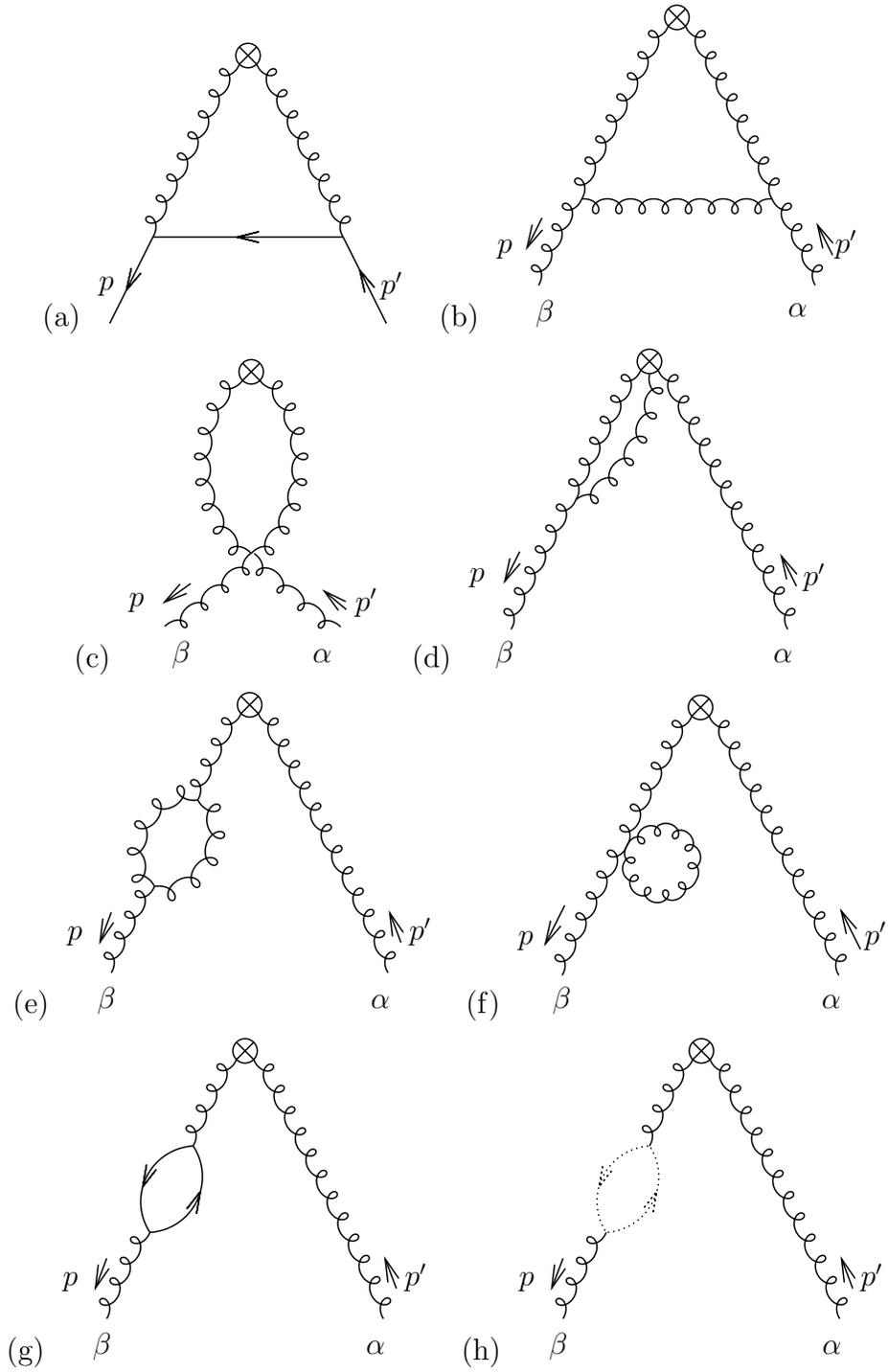
\begin{figure}
\leavevmode
\begin{center}
  (a)\hspace{0.1in}\input{diagg.pstex_t}\hspace{0.2in}
  (b)\hspace{0.1in}\input{diagh.pstex_t} \\ \vspace{0.1in}
  (c)\hspace{0.1in}\input{diagi.pstex_t}\hspace{0.2in}
  (d)\hspace{0.1in}\input{diagj.pstex_t} \\ \vspace{0.1in}
  (e)\hspace{0.1in}\input{diagk.pstex_t}\hspace{0.2in}
  (f)\hspace{0.1in}\input{diagl.pstex_t}  \\ \vspace{0.1in}
  (g)\hspace{0.1in}\input{diagm.pstex_t}\hspace{0.2in}
  (h)\hspace{0.1in}\input{diagn.pstex_t}
\caption{One-loop diagrams}
\label{diags2}
\end{center}
\end{figure}

To evaluate the UV divergent pieces to one loop, we insert the operators
$O_{1 \rightarrow 4}$ into Green functions with either quark/anti-quark legs
or two gluon legs, and calculate all the diagrams in
figs.~\ref{diags1} and~\ref{diags2}.
The bare vertices, symbolised by $\otimes$, have the following Feynman rules,
which also give the tree-level diagrams:
\bea
  O_1^{\mu\nu\lambda}: & & \raisebox{-0.5in}{\input{tree1.pstex_t}} \phantom{******}
   \frac{1}{2}\epsilon^{\mu\nu\lambda\sigma}\gamma_{\sigma}\gamma^5\\
  O_2^{\mu\nu}: & & \raisebox{-0.5in}{\input{tree1.pstex_t}} \phantom{******}
   \gamma^\mu P^\nu  \\
   & \& & \raisebox{-0.5in}{\input{tree2.pstex_t}} \phantom{******} g \gamma^\mu g^\nu_\alpha t^a \\
  O_4^{\mu\nu}: & & \raisebox{-0.5in}{\input{tree1.pstex_t}} \phantom{******}
   \frac{1}{2} \gamma^{\{\mu} P^{\nu\}}  \\
  & \& & \raisebox{-0.5in}{\input{tree2.pstex_t}} \phantom{******}
   \frac{1}{2} \gamma^{\{\mu} g^{\nu\}}_\alpha t^a  \\
  O_3^{\mu\nu}: & & \raisebox{-0.5in}{\input{tree3.pstex_t}} \phantom{******}
   \begin{array}{r}
   -p^{\{\mu} {\pp}^{\nu\}} g_{\alpha\beta} - p\cdot \pp g^\mu_{\{\alpha}
   g^\nu_{\beta\}} \\
  + p^{\{\mu} g^{\nu\}}_\alpha \pp_\beta\
   + {\pp}^{\{\mu} g^{\nu\}}_\beta p_\alpha \end{array} \\
  & \& & \raisebox{-0.5in}{\input{tree4.pstex_t}} \phantom{******}
   \begin{array}{r}
   -i g f^{abc} \left[ p^{\{\mu} g^{\nu\}[\gamma}g^{\beta]\alpha}
   - g^{\alpha\{\mu} g^{\nu\}[\gamma} p^{\beta]} \right.  \\
  q^{\{\mu} g^{\nu\}[\alpha}g^{\gamma]\beta}
   - g^{\beta\{\mu} g^{\nu\}[\alpha} q^{\gamma]}  \\
  \left. r^{\{\mu} g^{\nu\}[\beta}g^{\alpha]\gamma}
   - g^{\gamma\{\mu} g^{\nu\}[\beta} r^{\alpha]} \right] \end{array}
\eea
where $P = \frac{1}{2}(p + \pp)$, $\Delta = p - \pp$.

We evaluate the $\frac{1}{\epsilon}$ pole pieces of the Feynman diagrams
and keep only terms of order $\Delta$. For example, diagram~\ref{diags1}a
with $\otimes = O_2$ gives us, in the Feynman gauge:
\bea
  \lefteqn{C_F \pole\coupling\left[ -\frac{1}{3} \gamma^{\{\mu}P^{\nu\}}
   + \gamma^\mu P^\nu - \frac{1}{3} g^{\mu\nu} \slsh{P}
   - \frac{1}{2} i\Delta_\alpha \epsilon^{\mu\nu\alpha\sigma}\gamma_\sigma
   \gamma^5 \right]}& & \nonumber \\
  & = & C_F\pole\coupling\left[-\frac{2}{3} \mat{q}{O^{\mu\nu}_4}{q}_{\rm tree}
   + \mat{q}{O^{\mu\nu}_2}{q}_{\rm tree} + \frac{1}{3} g^{\mu\nu}
   \mat{q}{E_1}{q}_{\rm tree} -
   \mat{q}{\partial_\alpha O_1^{\mu\nu\alpha}}{q}_{\rm tree}\right].\nonumber\\
\eea
Adding up the diagrams in figs.~\ref{diags1} and~\ref{diags2}
and their left-right mirror images,
where appropriate, we get, neglecting $O(\epsilon^0)$, $O(\Delta^2)$ and
$g^{\mu\nu}$ terms,
\bea
  \mat{q}{O_1^{\mu\nu\lambda}}{q}_B & = & \left( 1 - C_F\pole\coupling \right)
   \mat{q}{O_1^{\mu\nu\lambda}}{q}_{\rm tree}  \\
  \mat{q}{O_2^{\mu\nu}}{q}_B & = & -\frac{8}{3}C_F\pole\coupling
   \mat{q}{O_4^{\mu\nu}}{q}_{\rm tree} \nonumber \\
   & & + \left(1 - C_F\pole\coupling\right)\mat{q}{O_2^{\mu\nu}}{q}_{\rm tree}
    \\
  \mat{q}{O_3^{\mu\nu}}{q}_B & = & \frac{8}{3}C_F\pole\coupling
   \mat{q}{O_4^{\mu\nu}}{q}_{\rm tree}\\
  \mat{q}{O_4^{\mu\nu}}{q}_B & = & \left(1 - \frac{11}{3}C_F\pole\coupling
   \right) \mat{q}{O_4^{\mu\nu}}{q}_{\rm tree}\\
  \mat{g}{O_1^{\mu\nu\lambda}}{g}_B & = & {\rm finite} \\
  \mat{g}{O_2^{\mu\nu}}{g}_B & = & \frac{2}{3}n_f\pole\coupling
   \mat{g}{O_3^{\mu\nu}}{g}_{\rm tree} \\
  \mat{g}{O_3^{\mu\nu}}{g}_B & = & \left(1 + \left(\frac{5}{3}
   - \frac{4}{3}n_f\right) \pole\coupling \right)
   \mat{g}{O_3^{\mu\nu}}{g}_{\rm tree} \\
  \mat{g}{O_4^{\mu\nu}}{g}_B & = & \mat{g}{O_2^{\mu\nu}}{g}_B.
\eea
Some of these poles come from the field-strength divergences of
figs.~\ref{diags1}b and~\ref{diags2}e-h
and are absorbed by multiplying each quark and gluon external leg by
$Z^{1/2}_\psi$ and $Z^{1/2}_A$ respectively, with
\bea
  Z_A & = & \left( 1 - \left(\frac{5}{3}C_F - \frac{2}{3}n_f\right)
   \pole\coupling\right), \nonumber \\
  Z_\psi & = & \left( 1 + C_F \pole \coupling \right).
\eea
Writing
\bea
  \mat{q}{O_B}{q}_R & = & Z_\psi \mat{q}{O_B}{q}_B, \nonumber \\
  \mat{g}{O_B}{g}_R & = & Z_A \mat{g}{O_B}{g}_B,
\eea
we have
\bea
  \mat{q}{O_{1,B}}{q}_R & = & \mat{q}{O_{1,B}}{q}_{\rm tree} \\
  \mat{q}{O_{2,B}}{q}_R & = & \mat{q}{O_{2,B}}{q}_{\rm tree}
   - \frac{8}{3}C_F\pole\coupling \mat{q}{O_{4,B}}{q}_{\rm tree} \\
  \mat{q}{O_{3,B}}{q}_R & = & \frac{8}{3}C_F\pole\coupling
   \mat{q}{O_{4,B}}{q}_{\rm tree} \\
  \mat{q}{O_{4,B}}{q}_R & = & \left(1 - \frac{8}{3}C_F\pole\coupling\right)
   \mat{q}{O_{4,B}}{q}_{\rm tree} \\
  \mat{g}{O_{1,B}}{g}_R & = & {\rm finite}\\
  \mat{g}{O_{2,B}}{g}_R & = & \frac{2}{3}n_f\pole\coupling
   \mat{g}{O_{3,B}}{g}_{\rm tree} \\
  \mat{g}{O_{3,B}}{g}_R & = & \left(1 - \frac{2}{3}n_f\pole\coupling\right)
   \mat{g}{O_{3,B}}{g}_{\rm tree} \\
  \mat{g}{O_{4,B}}{g}_R & = & \mat{g}{O_{2,B}}{g}_R.
\eea
The renormalisation matrix required to remove these remaining poles is then
given by eq.~(\ref{mix_op}).


\subsection*{Acknowledgements}

We should like to thank R. D. Ball and D. de Florian
for helpful discussions. This work is funded by PPARC
grant GR/L56374 in the U.K. B.E.W. is also very grateful to
C. Allton and S. Hands for additional support and encouragement, and
to PPARC for a research studentship.


\end{document}

%% file: vert1.pstex_t
\begin{picture}(0,0)%
\includegraphics{vert1.pstex}%
\end{picture}%
\setlength{\unitlength}{1973sp}%
\begingroup\makeatletter\ifx\SetFigFont\undefined%
\gdef\SetFigFont#1#2#3#4#5{%
  \reset@font\fontsize{#1}{#2pt}%
  \fontfamily{#3}\fontseries{#4}\fontshape{#5}%
  \selectfont}%
\fi\endgroup%
\begin{picture}(3816,2451)(1693,-3817)
\put(4201,-2461){\makebox(0,0)[lb]{\smash{\SetFigFont{10}{12.0}{\familydefault}{\mddefault}{\updefault}\special{ps: gsave 0 0 0 setrgbcolor}$k + \frac{1}{2}\Delta$\special{ps: grestore}}}}
\put(2026,-3736){\makebox(0,0)[lb]{\smash{\SetFigFont{10}{12.0}{\familydefault}{\mddefault}{\updefault}\special{ps: gsave 0 0 0 setrgbcolor}$\pp$\special{ps: grestore}}}}
\put(5101,-3736){\makebox(0,0)[lb]{\smash{\SetFigFont{10}{12.0}{\familydefault}{\mddefault}{\updefault}\special{ps: gsave 0 0 0 setrgbcolor}$p$\special{ps: grestore}}}}
\put(3826,-1636){\makebox(0,0)[lb]{\smash{\SetFigFont{10}{12.0}{\familydefault}{\mddefault}{\updefault}\special{ps: gsave 0 0 0 setrgbcolor}$\Gamma^{\lambda]\mu}(k)$\special{ps: grestore}}}}
\put(1876,-2461){\makebox(0,0)[lb]{\smash{\SetFigFont{10}{12.0}{\familydefault}{\mddefault}{\updefault}\special{ps: gsave 0 0 0 setrgbcolor}$k - \frac{1}{2}\Delta$\special{ps: grestore}}}}
\end{picture}

%% file: vert2.pstex_t
\begin{picture}(0,0)%
\includegraphics{vert2.pstex}%
\end{picture}%
\setlength{\unitlength}{1973sp}%
\begingroup\makeatletter\ifx\SetFigFont\undefined%
\gdef\SetFigFont#1#2#3#4#5{%
  \reset@font\fontsize{#1}{#2pt}%
  \fontfamily{#3}\fontseries{#4}\fontshape{#5}%
  \selectfont}%
\fi\endgroup%
\begin{picture}(3816,2901)(1693,-3817)
\put(2026,-3736){\makebox(0,0)[lb]{\smash{\SetFigFont{10}{12.0}{\familydefault}{\mddefault}{\updefault}\special{ps: gsave 0 0 0 setrgbcolor}$\pp$\special{ps: grestore}}}}
\put(5101,-3736){\makebox(0,0)[lb]{\smash{\SetFigFont{10}{12.0}{\familydefault}{\mddefault}{\updefault}\special{ps: gsave 0 0 0 setrgbcolor}$p$\special{ps: grestore}}}}
\put(4201,-2686){\makebox(0,0)[lb]{\smash{\SetFigFont{10}{12.0}{\familydefault}{\mddefault}{\updefault}\special{ps: gsave 0 0 0 setrgbcolor}$k + \frac{1}{2}\Delta$\special{ps: grestore}}}}
\put(1951,-2686){\makebox(0,0)[lb]{\smash{\SetFigFont{10}{12.0}{\familydefault}{\mddefault}{\updefault}\special{ps: gsave 0 0 0 setrgbcolor}$k - \frac{1}{2}\Delta$\special{ps: grestore}}}}
\put(3901,-1786){\makebox(0,0)[lb]{\smash{\SetFigFont{10}{12.0}{\familydefault}{\mddefault}{\updefault}\special{ps: gsave 0 0 0 setrgbcolor}$l + \frac{1}{2}\Delta$\special{ps: grestore}}}}
\put(2251,-1786){\makebox(0,0)[lb]{\smash{\SetFigFont{10}{12.0}{\familydefault}{\mddefault}{\updefault}\special{ps: gsave 0 0 0 setrgbcolor}$l - \frac{1}{2}\Delta$\special{ps: grestore}}}}
\put(3751,-1186){\makebox(0,0)[lb]{\smash{\SetFigFont{10}{12.0}{\familydefault}{\mddefault}{\updefault}\special{ps: gsave 0 0 0 setrgbcolor}$\Gamma^{\lambda]\mu}(l)$\special{ps: grestore}}}}
\end{picture}

%% file: vert3.pstex_t
\begin{picture}(0,0)%
\includegraphics{vert3.pstex}%
\end{picture}%
\setlength{\unitlength}{1973sp}%
\begingroup\makeatletter\ifx\SetFigFont\undefined%
\gdef\SetFigFont#1#2#3#4#5{%
  \reset@font\fontsize{#1}{#2pt}%
  \fontfamily{#3}\fontseries{#4}\fontshape{#5}%
  \selectfont}%
\fi\endgroup%
\begin{picture}(3816,2397)(1693,-3817)
\put(4201,-2461){\makebox(0,0)[lb]{\smash{\SetFigFont{10}{12.0}{\familydefault}{\mddefault}{\updefault}\special{ps: gsave 0 0 0 setrgbcolor}$k + \frac{1}{2}\Delta$\special{ps: grestore}}}}
\put(2026,-3736){\makebox(0,0)[lb]{\smash{\SetFigFont{10}{12.0}{\familydefault}{\mddefault}{\updefault}\special{ps: gsave 0 0 0 setrgbcolor}$\pp$\special{ps: grestore}}}}
\put(5101,-3736){\makebox(0,0)[lb]{\smash{\SetFigFont{10}{12.0}{\familydefault}{\mddefault}{\updefault}\special{ps: gsave 0 0 0 setrgbcolor}$p$\special{ps: grestore}}}}
\put(1876,-2461){\makebox(0,0)[lb]{\smash{\SetFigFont{10}{12.0}{\familydefault}{\mddefault}{\updefault}\special{ps: gsave 0 0 0 setrgbcolor}$k - \frac{1}{2}\Delta$\special{ps: grestore}}}}
\put(3751,-1636){\makebox(0,0)[lb]{\smash{\SetFigFont{10}{12.0}{\familydefault}{\mddefault}{\updefault}\special{ps: gsave 0 0 0 setrgbcolor}$\Gamma^{\lambda]\mu}(k,\Delta,\epsilon)$\special{ps: grestore}}}}
\end{picture}

%% file: exclusive.pstex_t
\begin{picture}(0,0)%
\includegraphics{exclusive.pstex}%
\end{picture}%
\setlength{\unitlength}{0.00050000in}%
\begingroup\makeatletter\ifx\SetFigFont\undefined%
\gdef\SetFigFont#1#2#3#4#5{%
  \reset@font\fontsize{#1}{#2pt}%
  \fontfamily{#3}\fontseries{#4}\fontshape{#5}%
  \selectfont}%
\fi\endgroup%
\begin{picture}(3547,2955)(2026,-3400)
\put(5401,-1111){\makebox(0,0)[lb]{\smash{\SetFigFont{12}{14.4}{\familydefault}{\mddefault}{\updefault}X}}}
\put(3226,-661){\makebox(0,0)[lb]{\smash{\SetFigFont{12}{14.4}{\familydefault}{\mddefault}{\updefault}$q$}}}
\put(2926,-3361){\makebox(0,0)[lb]{\smash{\SetFigFont{12}{14.4}{\familydefault}{\mddefault}{\updefault}p}}}
\put(5251,-3361){\makebox(0,0)[lb]{\smash{\SetFigFont{12}{14.4}{\familydefault}{\mddefault}{\updefault}p}}}
\put(2026,-2836){\makebox(0,0)[lb]{\smash{\SetFigFont{12}{14.4}{\familydefault}{\mddefault}{\updefault}$P-\frac{1}{2}\Delta$}}}
\put(5326,-2836){\makebox(0,0)[lb]{\smash{\SetFigFont{12}{14.4}{\familydefault}{\mddefault}{\updefault}$P+\frac{1}{2}\Delta$}}}
\put(2626,-1111){\makebox(0,0)[lb]{\smash{\SetFigFont{12}{14.4}{\familydefault}{\mddefault}{\updefault}$\gamma$}}}
\end{picture}

%% file: tree1.pstex_t
\begin{picture}(0,0)%
\includegraphics{tree1.pstex}%
\end{picture}%
\setlength{\unitlength}{0.00043700in}%
\begingroup\makeatletter\ifx\SetFigFont\undefined%
\gdef\SetFigFont#1#2#3#4#5{%
  \reset@font\fontsize{#1}{#2pt}%
  \fontfamily{#3}\fontseries{#4}\fontshape{#5}%
  \selectfont}%
\fi\endgroup%
\begin{picture}(2024,2054)(249,-1433)
\put(361,-691){\makebox(0,0)[lb]{\smash{\SetFigFont{10}{12.0}{\familydefault}{\mddefault}{\updefault}$p$}}}
\put(2071,-691){\makebox(0,0)[lb]{\smash{\SetFigFont{10}{12.0}{\familydefault}{\mddefault}{\updefault}$\pp$}}}
\end{picture}

%% file: tree2.pstex_t
\begin{picture}(0,0)%
\includegraphics{tree2.pstex}%
\end{picture}%
\setlength{\unitlength}{0.00043700in}%
\begingroup\makeatletter\ifx\SetFigFont\undefined%
\gdef\SetFigFont#1#2#3#4#5{%
  \reset@font\fontsize{#1}{#2pt}%
  \fontfamily{#3}\fontseries{#4}\fontshape{#5}%
  \selectfont}%
\fi\endgroup%
\begin{picture}(2024,2446)(249,-1825)
\put(1126,-1771){\makebox(0,0)[lb]{\smash{\SetFigFont{10}{12.0}{\familydefault}{\mddefault}{\updefault}$\alpha,a$}}}
\end{picture}

%% file: tree3.pstex_t
\begin{picture}(0,0)%
\includegraphics{tree3.pstex}%
\end{picture}%
\setlength{\unitlength}{0.00043700in}%
\begingroup\makeatletter\ifx\SetFigFont\undefined%
\gdef\SetFigFont#1#2#3#4#5{%
  \reset@font\fontsize{#1}{#2pt}%
  \fontfamily{#3}\fontseries{#4}\fontshape{#5}%
  \selectfont}%
\fi\endgroup%
\begin{picture}(2610,2131)(631,-1600)
\put(631,-1546){\makebox(0,0)[lb]{\smash{\SetFigFont{10}{12.0}{\familydefault}{\mddefault}{\updefault}$\beta$}}}
\put(3106,-736){\makebox(0,0)[lb]{\smash{\SetFigFont{10}{12.0}{\familydefault}{\mddefault}{\updefault}$\pp$}}}
\put(676,-736){\makebox(0,0)[lb]{\smash{\SetFigFont{10}{12.0}{\familydefault}{\mddefault}{\updefault}$p$}}}
\put(3241,-1546){\makebox(0,0)[lb]{\smash{\SetFigFont{10}{12.0}{\familydefault}{\mddefault}{\updefault}$\alpha$}}}
\end{picture}

%% file: tree4.pstex_t
\begin{picture}(0,0)%
\includegraphics{tree4.pstex}%
\end{picture}%
\setlength{\unitlength}{0.00043700in}%
\begingroup\makeatletter\ifx\SetFigFont\undefined%
\gdef\SetFigFont#1#2#3#4#5{%
  \reset@font\fontsize{#1}{#2pt}%
  \fontfamily{#3}\fontseries{#4}\fontshape{#5}%
  \selectfont}%
\fi\endgroup%
\begin{picture}(2655,2401)(406,-1780)
\put(541,-1726){\makebox(0,0)[lb]{\smash{\SetFigFont{10}{12.0}{\familydefault}{\mddefault}{\updefault}$\gamma,c$}}}
\put(3061,-1006){\makebox(0,0)[lb]{\smash{\SetFigFont{10}{12.0}{\familydefault}{\mddefault}{\updefault}$p$}}}
\put(2071,-1051){\makebox(0,0)[lb]{\smash{\SetFigFont{10}{12.0}{\familydefault}{\mddefault}{\updefault}$q$}}}
\put(406,-1006){\makebox(0,0)[lb]{\smash{\SetFigFont{10}{12.0}{\familydefault}{\mddefault}{\updefault}$r$}}}
\put(2836,-1726){\makebox(0,0)[lb]{\smash{\SetFigFont{10}{12.0}{\familydefault}{\mddefault}{\updefault}$\alpha,a$}}}
\put(1666,-1726){\makebox(0,0)[lb]{\smash{\SetFigFont{10}{12.0}{\familydefault}{\mddefault}{\updefault}$\beta,b$}}}
\end{picture}

%% file: diaga.pstex_t
\begin{picture}(0,0)%
\includegraphics{diaga.pstex}%
\end{picture}%
\setlength{\unitlength}{0.00052500in}%
\begingroup\makeatletter\ifx\SetFigFont\undefined%
\gdef\SetFigFont#1#2#3#4#5{%
  \reset@font\fontsize{#1}{#2pt}%
  \fontfamily{#3}\fontseries{#4}\fontshape{#5}%
  \selectfont}%
\fi\endgroup%
\begin{picture}(3105,2954)(991,-2333)
\put(4096,-1996){\makebox(0,0)[lb]{\smash{\SetFigFont{12}{14.4}{\familydefault}{\mddefault}{\updefault}$\pp$}}}
\put(991,-1996){\makebox(0,0)[lb]{\smash{\SetFigFont{12}{14.4}{\familydefault}{\mddefault}{\updefault}$p$}}}
\end{picture}

%% file: diagb.pstex_t
\begin{picture}(0,0)%
\includegraphics{diagb.pstex}%
\end{picture}%
\setlength{\unitlength}{0.00052500in}%
\begingroup\makeatletter\ifx\SetFigFont\undefined%
\gdef\SetFigFont#1#2#3#4#5{%
  \reset@font\fontsize{#1}{#2pt}%
  \fontfamily{#3}\fontseries{#4}\fontshape{#5}%
  \selectfont}%
\fi\endgroup%
\begin{picture}(2992,2954)(1081,-2333)
\put(4006,-1951){\makebox(0,0)[lb]{\smash{\SetFigFont{12}{14.4}{\familydefault}{\mddefault}{\updefault}$\pp$}}}
\put(1081,-1951){\makebox(0,0)[lb]{\smash{\SetFigFont{12}{14.4}{\familydefault}{\mddefault}{\updefault}$p$}}}
\end{picture}

%% file: diagc.pstex_t
\begin{picture}(0,0)%
\includegraphics{diagc.pstex}%
\end{picture}%
\setlength{\unitlength}{0.00052500in}%
\begingroup\makeatletter\ifx\SetFigFont\undefined%
\gdef\SetFigFont#1#2#3#4#5{%
  \reset@font\fontsize{#1}{#2pt}%
  \fontfamily{#3}\fontseries{#4}\fontshape{#5}%
  \selectfont}%
\fi\endgroup%
\begin{picture}(3645,3346)(721,-2725)
\put(721,-1951){\makebox(0,0)[lb]{\smash{\SetFigFont{12}{14.4}{\familydefault}{\mddefault}{\updefault}$p$}}}
\put(3961,-2671){\makebox(0,0)[lb]{\smash{\SetFigFont{12}{14.4}{\familydefault}{\mddefault}{\updefault}$\alpha$}}}
\put(991,-2671){\makebox(0,0)[lb]{\smash{\SetFigFont{12}{14.4}{\familydefault}{\mddefault}{\updefault}$\beta$}}}
\put(4366,-1996){\makebox(0,0)[lb]{\smash{\SetFigFont{12}{14.4}{\familydefault}{\mddefault}{\updefault}$\pp$}}}
\end{picture}

%% file: diagd.pstex_t
\begin{picture}(0,0)%
\includegraphics{diagd.pstex}%
\end{picture}%
\setlength{\unitlength}{0.00052500in}%
\begingroup\makeatletter\ifx\SetFigFont\undefined%
\gdef\SetFigFont#1#2#3#4#5{%
  \reset@font\fontsize{#1}{#2pt}%
  \fontfamily{#3}\fontseries{#4}\fontshape{#5}%
  \selectfont}%
\fi\endgroup%
\begin{picture}(2985,3391)(1756,-4120)
\put(4546,-4066){\makebox(0,0)[lb]{\smash{\SetFigFont{12}{14.4}{\familydefault}{\mddefault}{\updefault}$\alpha$}}}
\put(1756,-3346){\makebox(0,0)[lb]{\smash{\SetFigFont{12}{14.4}{\familydefault}{\mddefault}{\updefault}$p$}}}
\put(4681,-3346){\makebox(0,0)[lb]{\smash{\SetFigFont{12}{14.4}{\familydefault}{\mddefault}{\updefault}$\pp$}}}
\put(1801,-4066){\makebox(0,0)[lb]{\smash{\SetFigFont{12}{14.4}{\familydefault}{\mddefault}{\updefault}$\beta$}}}
\end{picture}

%% file: diage.pstex_t
\begin{picture}(0,0)%
\includegraphics{diage.pstex}%
\end{picture}%
\setlength{\unitlength}{0.00052500in}%
\begingroup\makeatletter\ifx\SetFigFont\undefined%
\gdef\SetFigFont#1#2#3#4#5{%
  \reset@font\fontsize{#1}{#2pt}%
  \fontfamily{#3}\fontseries{#4}\fontshape{#5}%
  \selectfont}%
\fi\endgroup%
\begin{picture}(2924,2954)(1149,-2333)
\put(1171,-1681){\makebox(0,0)[lb]{\smash{\SetFigFont{12}{14.4}{\familydefault}{\mddefault}{\updefault}$p$}}}
\put(3916,-1681){\makebox(0,0)[lb]{\smash{\SetFigFont{12}{14.4}{\familydefault}{\mddefault}{\updefault}$\pp$}}}
\end{picture}

%% file: diagf.pstex_t
\begin{picture}(0,0)%
\includegraphics{diagf.pstex}%
\end{picture}%
\setlength{\unitlength}{0.00052500in}%
\begingroup\makeatletter\ifx\SetFigFont\undefined%
\gdef\SetFigFont#1#2#3#4#5{%
  \reset@font\fontsize{#1}{#2pt}%
  \fontfamily{#3}\fontseries{#4}\fontshape{#5}%
  \selectfont}%
\fi\endgroup%
\begin{picture}(3285,3346)(901,-2725)
\put(901,-1726){\makebox(0,0)[lb]{\smash{\SetFigFont{12}{14.4}{\familydefault}{\mddefault}{\updefault}$p$}}}
\put(3916,-2671){\makebox(0,0)[lb]{\smash{\SetFigFont{12}{14.4}{\familydefault}{\mddefault}{\updefault}$\alpha$}}}
\put(1036,-2671){\makebox(0,0)[lb]{\smash{\SetFigFont{12}{14.4}{\familydefault}{\mddefault}{\updefault}$\beta$}}}
\put(4186,-1681){\makebox(0,0)[lb]{\smash{\SetFigFont{12}{14.4}{\familydefault}{\mddefault}{\updefault}$\pp$}}}
\end{picture}

%% file: diagg.pstex_t
\begin{picture}(0,0)%
\includegraphics{diagg.pstex}%
\end{picture}%
\setlength{\unitlength}{0.00052500in}%
\begingroup\makeatletter\ifx\SetFigFont\undefined%
\gdef\SetFigFont#1#2#3#4#5{%
  \reset@font\fontsize{#1}{#2pt}%
  \fontfamily{#3}\fontseries{#4}\fontshape{#5}%
  \selectfont}%
\fi\endgroup%
\begin{picture}(2992,2954)(1081,-2333)
\put(1081,-1951){\makebox(0,0)[lb]{\smash{\SetFigFont{12}{14.4}{\familydefault}{\mddefault}{\updefault}$p$}}}
\put(4051,-1996){\makebox(0,0)[lb]{\smash{\SetFigFont{12}{14.4}{\familydefault}{\mddefault}{\updefault}$\pp$}}}
\end{picture}

%% file: diagh.pstex_t
\begin{picture}(0,0)%
\includegraphics{diagh.pstex}%
\end{picture}%
\setlength{\unitlength}{0.00052500in}%
\begingroup\makeatletter\ifx\SetFigFont\undefined%
\gdef\SetFigFont#1#2#3#4#5{%
  \reset@font\fontsize{#1}{#2pt}%
  \fontfamily{#3}\fontseries{#4}\fontshape{#5}%
  \selectfont}%
\fi\endgroup%
\begin{picture}(3555,3340)(766,-2719)
\put(766,-1951){\makebox(0,0)[lb]{\smash{\SetFigFont{12}{14.4}{\familydefault}{\mddefault}{\updefault}$p$}}}
\put(3826,-2626){\makebox(0,0)[lb]{\smash{\SetFigFont{12}{14.4}{\familydefault}{\mddefault}{\updefault}$\alpha$}}}
\put(1171,-2671){\makebox(0,0)[lb]{\smash{\SetFigFont{12}{14.4}{\familydefault}{\mddefault}{\updefault}$\beta$}}}
\put(4321,-1951){\makebox(0,0)[lb]{\smash{\SetFigFont{12}{14.4}{\familydefault}{\mddefault}{\updefault}$\pp$}}}
\end{picture}

%% file: diagi.pstex_t
\begin{picture}(0,0)%
\includegraphics{diagi.pstex}%
\end{picture}%
\setlength{\unitlength}{0.00052500in}%
\begingroup\makeatletter\ifx\SetFigFont\undefined%
\gdef\SetFigFont#1#2#3#4#5{%
  \reset@font\fontsize{#1}{#2pt}%
  \fontfamily{#3}\fontseries{#4}\fontshape{#5}%
  \selectfont}%
\fi\endgroup%
\begin{picture}(2385,3211)(91,-2725)
\put(2476,-2131){\makebox(0,0)[lb]{\smash{\SetFigFont{12}{14.4}{\familydefault}{\mddefault}{\updefault}$\pp$}}}
\put(2026,-2671){\makebox(0,0)[lb]{\smash{\SetFigFont{12}{14.4}{\familydefault}{\mddefault}{\updefault}$\alpha$}}}
\put(541,-2671){\makebox(0,0)[lb]{\smash{\SetFigFont{12}{14.4}{\familydefault}{\mddefault}{\updefault}$\beta$}}}
\put( 91,-2086){\makebox(0,0)[lb]{\smash{\SetFigFont{12}{14.4}{\familydefault}{\mddefault}{\updefault}$p$}}}
\end{picture}

%% file: diagj.pstex_t
\begin{picture}(0,0)%
\includegraphics{diagj.pstex}%
\end{picture}%
\setlength{\unitlength}{0.00052500in}%
\begingroup\makeatletter\ifx\SetFigFont\undefined%
\gdef\SetFigFont#1#2#3#4#5{%
  \reset@font\fontsize{#1}{#2pt}%
  \fontfamily{#3}\fontseries{#4}\fontshape{#5}%
  \selectfont}%
\fi\endgroup%
\begin{picture}(3510,3301)(766,-2680)
\put(3961,-2626){\makebox(0,0)[lb]{\smash{\SetFigFont{12}{14.4}{\familydefault}{\mddefault}{\updefault}$\alpha$}}}
\put(4276,-1816){\makebox(0,0)[lb]{\smash{\SetFigFont{12}{14.4}{\familydefault}{\mddefault}{\updefault}$\pp$}}}
\put(766,-1771){\makebox(0,0)[lb]{\smash{\SetFigFont{12}{14.4}{\familydefault}{\mddefault}{\updefault}$p$}}}
\put(1036,-2626){\makebox(0,0)[lb]{\smash{\SetFigFont{12}{14.4}{\familydefault}{\mddefault}{\updefault}$\beta$}}}
\end{picture}

%% file: diagk.pstex_t
\begin{picture}(0,0)%
\includegraphics{diagk.pstex}%
\end{picture}%
\setlength{\unitlength}{0.00052500in}%
\begingroup\makeatletter\ifx\SetFigFont\undefined%
\gdef\SetFigFont#1#2#3#4#5{%
  \reset@font\fontsize{#1}{#2pt}%
  \fontfamily{#3}\fontseries{#4}\fontshape{#5}%
  \selectfont}%
\fi\endgroup%
\begin{picture}(3600,3346)(721,-2725)
\put(721,-1951){\makebox(0,0)[lb]{\smash{\SetFigFont{12}{14.4}{\familydefault}{\mddefault}{\updefault}$p$}}}
\put(3916,-2671){\makebox(0,0)[lb]{\smash{\SetFigFont{12}{14.4}{\familydefault}{\mddefault}{\updefault}$\alpha$}}}
\put(1036,-2671){\makebox(0,0)[lb]{\smash{\SetFigFont{12}{14.4}{\familydefault}{\mddefault}{\updefault}$\beta$}}}
\put(4321,-1951){\makebox(0,0)[lb]{\smash{\SetFigFont{12}{14.4}{\familydefault}{\mddefault}{\updefault}$\pp$}}}
\end{picture}

%% file: diagl.pstex_t
\begin{picture}(0,0)%
\includegraphics{diagl.pstex}%
\end{picture}%
\setlength{\unitlength}{0.00052500in}%
\begingroup\makeatletter\ifx\SetFigFont\undefined%
\gdef\SetFigFont#1#2#3#4#5{%
  \reset@font\fontsize{#1}{#2pt}%
  \fontfamily{#3}\fontseries{#4}\fontshape{#5}%
  \selectfont}%
\fi\endgroup%
\begin{picture}(3645,3301)(721,-2680)
\put(4366,-1951){\makebox(0,0)[lb]{\smash{\SetFigFont{12}{14.4}{\familydefault}{\mddefault}{\updefault}$\pp$}}}
\put(1081,-2626){\makebox(0,0)[lb]{\smash{\SetFigFont{12}{14.4}{\familydefault}{\mddefault}{\updefault}$\beta$}}}
\put(3916,-2626){\makebox(0,0)[lb]{\smash{\SetFigFont{12}{14.4}{\familydefault}{\mddefault}{\updefault}$\alpha$}}}
\put(721,-1951){\makebox(0,0)[lb]{\smash{\SetFigFont{12}{14.4}{\familydefault}{\mddefault}{\updefault}$p$}}}
\end{picture}

%% file: diagm.pstex_t
\begin{picture}(0,0)%
\includegraphics{diagm.pstex}%
\end{picture}%
\setlength{\unitlength}{0.00052500in}%
\begingroup\makeatletter\ifx\SetFigFont\undefined%
\gdef\SetFigFont#1#2#3#4#5{%
  \reset@font\fontsize{#1}{#2pt}%
  \fontfamily{#3}\fontseries{#4}\fontshape{#5}%
  \selectfont}%
\fi\endgroup%
\begin{picture}(3600,3346)(721,-2725)
\put(721,-1951){\makebox(0,0)[lb]{\smash{\SetFigFont{12}{14.4}{\familydefault}{\mddefault}{\updefault}$p$}}}
\put(3916,-2671){\makebox(0,0)[lb]{\smash{\SetFigFont{12}{14.4}{\familydefault}{\mddefault}{\updefault}$\alpha$}}}
\put(1036,-2671){\makebox(0,0)[lb]{\smash{\SetFigFont{12}{14.4}{\familydefault}{\mddefault}{\updefault}$\beta$}}}
\put(4321,-1951){\makebox(0,0)[lb]{\smash{\SetFigFont{12}{14.4}{\familydefault}{\mddefault}{\updefault}$\pp$}}}
\end{picture}

%% file: diagn.pstex_t
\begin{picture}(0,0)%
\includegraphics{diagn.pstex}%
\end{picture}%
\setlength{\unitlength}{0.00052500in}%
\begingroup\makeatletter\ifx\SetFigFont\undefined%
\gdef\SetFigFont#1#2#3#4#5{%
  \reset@font\fontsize{#1}{#2pt}%
  \fontfamily{#3}\fontseries{#4}\fontshape{#5}%
  \selectfont}%
\fi\endgroup%
\begin{picture}(3600,3346)(721,-2725)
\put(721,-1951){\makebox(0,0)[lb]{\smash{\SetFigFont{12}{14.4}{\familydefault}{\mddefault}{\updefault}$p$}}}
\put(3916,-2671){\makebox(0,0)[lb]{\smash{\SetFigFont{12}{14.4}{\familydefault}{\mddefault}{\updefault}$\alpha$}}}
\put(1036,-2671){\makebox(0,0)[lb]{\smash{\SetFigFont{12}{14.4}{\familydefault}{\mddefault}{\updefault}$\beta$}}}
\put(4321,-1951){\makebox(0,0)[lb]{\smash{\SetFigFont{12}{14.4}{\familydefault}{\mddefault}{\updefault}$\pp$}}}
\end{picture}